\documentclass{AIAA}
\usepackage{subfigure}
\usepackage{epstopdf}
\usepackage[finalnew]{trackchanges} %[inline] shows all changes. [finalnew] accept all changes.
\usepackage{fancyhdr}
\renewcommand{\thispagestyle}[1]{} % do nothing
\begin{document}

\title{Modeling radio communication blackout and blackout mitigation in hypersonic vehicles}

\author{Madhusudhan Kundrapu\footnote{Associate Research Scientist,Tech-X Corporation, 5621 Arapahoe Avenue, Boulder, CO, 80303. Member AIAA.}, John Loverich\footnote{Research Scientist,Tech-X Corporation, 5621 Arapahoe Avenue, Boulder, CO, 80303.}, Kristian Beckwith\footnote{Research Scientist,Tech-X Corporation, 5621 Arapahoe Avenue, Boulder, CO, 80303.}, Peter Stoltz\footnote{Vice President,Tech-X Corporation, 5621 Arapahoe Avenue, Boulder, CO, 80303.}}
\affiliation{Tech-X Corporation, Boulder, CO, 80303, USA}
\author{Alexey Shashurin\footnote{Research Scientist, Mechanical and Aerospace Engineering, 801 22nd Street, NW, Washington, DC, 20052.} and Michael Keidar\footnote{Professor, Mechanical and Aerospace Engineering, 801 22nd Street, NW, Washington, DC, 20052. Associate Fellow AIAA.}}
\affiliation{The George Washington University, Washington, DC, 20052, USA}

\begin{abstract}
A procedure for the modeling and analysis of radio communication blackout of hypersonic vehicles \change[]{was}{is} presented. The weakly ionized plasma generated around the surface of a hypersonic reentry vehicle \change[]{was}{is} simulated using full Navier-Stokes equations in multi-species single fluid form. A seven species air chemistry model \change[]{was}{is} used to compute the individual species densities in air including ionization - plasma densities \change[]{were}{are} compared with experiment.  The electromagnetic wave's interaction with the plasma layer \change[]{was}{is} modeled using multi-fluid equations for fluid transport and full Maxwell's equations for the electromagnetic fields. The multi-fluid solver \change[]{was}{is} verified for a whistler wave propagating through a slab. First principles radio communication blackout over a hypersonic vehicle \change[]{was}{is} demonstrated along with a simple blackout mitigation scheme using a magnetic window.
\end{abstract}

\maketitle

\section*{Nomenclature}
\noindent\begin{tabular}{@{}lcl@{}}
\textit{A}  &=& amplitude of oscillation \\
\textit{$\vec{B}$} &=&    magnetic field, T\\
\textit{$\vec{B}$}$_{0}$ &=&    background static magnetic field, T\\
\textit{c}  &=& speed of light, m/s \\
\textit{c}$_{p}$&=& specific heat at constant pressure, J/kg K \\
\textit{c}$_{v}$&=& specific heat at constant volume, J/kg K \\
\textit{$\vec{E}$} &=& electric field, V/m\\
\textit{e} &=& total (internal+kinetic+chemical) energy of the fluid , J/m$^3$ \\
           &=& electron \\
\textit{f}   &=& number of degrees of freedom \\
           &=& frequency, Hz \\
\textit{H}  &=& enthalpy of formation, J/particle \\
\textit{I}  &=& identity matrix \\
\textit{k}$_{T}$  &=& thermal conductivity, W/m K \\
\textit{k} &=& wave number \\
\textit{k}$_{B}$  &=& Boltzmann constant, J/K \\
\textit{m}  &=& mass of a unit species, kg \\
\textit{M}  &=& molecular weight, g/mol \\
\textit{n} &=& \add[]{number density, 1/m$^{3}$} \\
\textit{N}  &=& number of species or fluids in the system \\
\textit{q}  &=& charge of a unit species, C \\
\textit{Q}  &=& internal energy exchange rate per unit volume, W/m$^3$ \\
\textit{Q}$_{EM}$  &=& energy density of the electromagnetic wave, J/m$^3$ \\
\textit{R} &=& gas constant, J/kg K \\
\textit{$\vec{R}$} &=& momentum exchange rate per unit volume, N/m$^3$\\ 
\textit{t} &=& time, s\\
\textit{T} &=& temperature, K \\ 
\textit{$\vec{u}$}  &=& velocity, m/s\\
\end{tabular} \\
\clearpage
\noindent\begin{tabular}{@{}lcl@{}}
\textit{$\gamma$}   &=& gas constant (\textit{c}$_{p}$/\textit{c}$_{v}$) \\
\textit{$\mu$}  &=& dynamic viscosity, N s/m$^{2}$ \\
                &=& reduced mass, kg \\
\textit{$\rho$}  &=& density, kg/m$^{3}$ \\
\textit{$\tau$}  &=& stress tensor, N/m$^{2}$ \\
\textit{$\sigma$}  &=& collision diameter, m \\
\textit{$\epsilon$}$_0$ &=& permittivity of free space, F/m \\
\textit{$\zeta$} &=& collision time, s \\
\textit{$\Omega$}  &=& collision integral \\
\textit{$\Omega_{ce}$} &=& electron cyclotron frequency, radians/s\\
\textit{$\Omega_{ci}$} &=& ion cyclotron frequency, radians/s\\
\textit{$\Omega_{pe}$} &=& electron plasma frequency, radians/s\\
\textit{$\Omega_{pi}$} &=& ion plasma frequency, radians/s\\
Subscripts \\
 
\textit{$\alpha$} &=& index of fluids \\
\textit{e} &=& \add[]{electron} \\
\textit{i} &=& index of species \remove[]{other than the species $\alpha$}\add[]{or fluids} \\
           &=& \add[]{ion} \\
Supercripts \\
 
\textit{T} &=& transpose \\
\end{tabular}

\section{Introduction}
Hypersonic vehicles are subjected to severe aerothermal heating due to the formation of shock waves in front of the vehicle. Flow Mach numbers exceeding four are classified as hypersonic\cite{bertin1994hypersonic}\add[]{. } \change[]{and in this regime the kinetic energy of the flow, when converted to internal energy through the shock provides a significant increase in the fluid temperature}{The shock wave converts the kinetic energy to internal energy and thereby increases the fluid temperatures significantly.} The temperatures quite often exceed the dissociation and ionization limits of the flow species and results in the formation of a weakly ionized plasma layer around the vehicle. The electrons in the plasma layer may interrupt the propagation of radio frequency electromagnetic waves\add[]{,} if the plasma electron oscillation frequency exceeds that of the electromagnetic wave frequency. This phenomenon is commonly called radio communication blackout. For instance, a 1.6 GHz radio wave will be interrupted by a plasma layer of density \remove[]{of} $3.5\times 10^{16}$ \add[]{m}$^{-3}$.   Blackout mitigation is an important requirement for the design of hypersonic vehicles, especially for those vehicles in steady state hypersonic flight such as those envisioned by NASA and the US Air Force. A few mitigation mechanisms described in the literature are the magnetic window\cite{hodara1961,manning2009analysis,thoma2009electromagnetic,starkey2003electromagnetic,usui2000Computer}, electrophilic fluid injection\cite{meyer2007system},  wave frequency modification, aerodynamic shape modification, $E\times B$ drift\cite{{gillman2010review},{keidar2008electromagnetic}}, resonant transmission\cite{sternberg2009resonant}, time varying magnetic field\cite{stenzel2013new} and electron acoustic wave transmission\cite{mudaliar2012radiation}. The magnetic window uses a static magnetic field to convert the free space radio wave to a whistler wave in the plasma.\cite{stenzel1976whistler,usui2000Computer} Electrophilic injection uses an electrophilic substance injected into the fluid to decrease the electron density.  Wave frequency and aerodynamic shape modification have design limitations so may be impractical in many cases.  The $E\times B$ drift accelerates the ions in the layer there by decreasing the plasma density near the antenna.  Resonant transmission uses surface wave resonance to enhance transmission through the plasma layer.  The time varying magnetic field approach uses the hall effect to expel ions.  Electron acoustic wave transmission works by converting the wave into an electron acoustic wave in the plasma layer.  Numerical simulations of blackout mitigation techniques are valuable during the design phase of hypersonic vehicles.

Radio communication blackout modeling of aerospace vehicles with full wave electromagnetics has been investigated by several groups with many different codes.  Takahashi\cite{{takahashi2014prediction},{takahashi2014examination}} used a CFD tool to compute the plasma distribution and then a FDTD solver with a modified permittivity to account for the presence of a plasma.  \add[]{Usui} \cite{usui2000Computer} \add[]{ demonstrated one dimensional PIC simulation of the whistler wave propagation in an assumed dense plasma profile.} Thoma\cite{thoma2009electromagnetic} used the high density FDTD PIC code LSP to investigate the magnetic window with a horn antenna surrounded by an assumed plasma distribution.  Visbal\cite{visbal2008high} used a multi-fluid electromagnetic approach to modeling radio communication blackout on an over-set mesh, without an investigation of steady magnetic field effects. \add[]{Starkey}\cite{starkey2003electromagnetic} \add[]{used one dimensional finite volume inviscid solver to estimate the plasma density on hypersonic vehicles along their trajectories and analyzed the whistler wave propagation using an imposed constant magnetic field.} 

\change[]{The scope of this paper is to show the modeling approach that works for realistic vehicles in complex geometries and can be used to simulate the feasibility of blackout mitigation devices for hypersonic vehicles}{The present paper shows the combined modeling approach to simulate the blackout mitigation devices and the surrounded reactive flows in complex geometries of realistic hypersonic vehicles}. \add[]{The finite volume simulation model used in the present work simulates the multi-species/fluid flow, reactions, plasma generation, and electromagnetic fields in the fluid without ignoring the plasma waves.} RAM C reentry vehicle is used for this demonstration as there is a significant experimental data for comparison.\cite{jones1972electrostatic}  USim\cite{{loverich2013nautilus},{shashurin2014laboratory},{kundrapu2013modeling}}, a commercial code developed by Tech-X Corporation for general fluid plasma modeling on unstructured grids, is used for all simulations in this paper. This paper is organized as follows: (1) Modeling and simulation of the multi-species hypersonic flow over the RAM C reentry vehicle to obtain the plasma density distribution (2) Validation of the plasma density distribution with the results from literature (3) Modeling of electromagnetic wave propagation into the plasma and validation with the dispersion relation and (4) Finally, the propagation of plane EM wave on to the vehicle's surface through the plasma layer using a magnetic window and the whistler wave conversion.

\section{Mathematical formulation}
\label{sec:mathematicalFormulation}
\subsection{Bulk fluid transport}\label{bulkFluid}
A generalized model for simulating the compressible flow with reacting multi-species is given in this section. The Navier-Stokes equations in conservative form Eqs. (\ref{eq:mass})--(\ref{eq:totalEnergy}) \change[]{were}{are} used for the conservation of fluid mass, momentum, and total energy respectively. The total energy $e$ in Eq. (\ref{eq:totalEnergy}) is the sum of internal energy, kinetic energy and \remove[]{the} chemical energy of the fluid. \add[]{The upper limit $N$ in the summation of Eqs.} (\ref{eq:avgDensity}) and (\ref{eq:totalEnergyDef}) \add[]{is the total number of species in the system.} The fluid \change[]{was}{is} assumed to be Newtonian and obeys the Stoke's hypothesis of zero bulk viscosity. Further, the fluid obeys the ideal gas law for equation of state.

\begin{equation}
\frac{\partial \rho}{\partial t} + \nabla\cdot\left(\rho \vec{u}\right) = 0
\label{eq:mass}
\end{equation}
\begin{equation}
\frac{\partial \left(\rho \vec{u}\right)}{\partial t} + \nabla\cdot\left(\rho \vec{u}\vec{u} + pI\right) = \nabla\cdot\mathbf{\tau}
\label{eq:momentum}
\end{equation}
\begin{equation}
\frac{\partial \left(e\right)}{\partial t} + \nabla\cdot\left(\vec{u}\left(e+p\right)\right) = \nabla\cdot\left(\tau \cdot \vec{u}\right) + \nabla\cdot\left(k_{T}\nabla T\right)
\label{eq:totalEnergy}
\end{equation}

where,
\begin{equation}
 \rho = \sum\limits_{i}^{N}{n_{i}m_{i}}
 \label{eq:avgDensity}
\end{equation}
\begin{equation}
  p = \rho R T
   \end{equation}
\begin{equation}
   \tau =  -\frac{2}{3}\mu\left(\nabla\cdot\vec{u}\right)I + \mu\left(\nabla\vec{u} + \left(\nabla\vec{u}\right)^{T} \right)
\end{equation}  
\begin{equation}   
   e = \frac{p}{\gamma -1} + \frac{1}{2}\rho \vec{u}\cdot\vec{u} +\sum\limits_{i}^{N}{n_{i}H_{i}}
   \label{eq:totalEnergyDef}
\end{equation}
 and 
\begin{equation}   
   \gamma = \frac{c_p}{c_p - R}
\end{equation}
   
\subsection{Species transport}
The mass conservation of the individual species in the bulk fluid is satisfied separately for each of the species using Eq. (\ref{eq:speciesCont}). The velocity $\vec{u}$ is the same as that of the bulk fluid. The right hand side of Eq. (\ref{eq:speciesCont}) represents the rate of change of species density due to the chemical reactions. \add[]{In a given chemical reaction, the rates of change of species density are obtained using the forward and backward reaction rates and the existing number densities of the reactants and products. The rates of change of a species $i$ in all of the reactions are added to get $s_i$.}

\begin{equation}
\frac{\partial n_i}{\partial t} + \nabla\cdot\left(\vec{u} n_i\right) = s_i
\label{eq:speciesCont}
\end{equation}

\subsection{Material properties}
The properties viscosity, thermal conductivity and \remove[]{the} specific heat of the individual species \change[]{were}{are} obtained from the kinetic theory of gases as given by the Eqs. (\ref{eq:viscosity})--(\ref{eq:specificHeat}). The fluid thermal conductivity $k$ and viscosity $\mu$ in Eqs. (\ref{eq:mass})--(\ref{eq:totalEnergy}) \change[]{were}{are} obtained using mole fraction averaging while the specific heat $c_p$ \change[]{was}{is} obtained using the mass fraction averaging. The gas constant $R$ \change[]{was}{is} computed using the mole fraction averaged molecular weight.

\begin{equation}
{c_{p}}_{i} = \left(\frac{f}{2}+1\right)R_{i}
\label{eq:specificHeatp}
\end{equation}
\begin{equation}
\mu_{i} = \frac{5}{16}\frac{\sqrt{\pi m_{i}k_{B}T}}{\left(\pi \sigma^2 \Omega\right)}
\label{eq:viscosity}
\end{equation}
\begin{equation}
k_{i} = \frac{5}{2}{c_{v}}_{i}\mu_{i}
\label{eq:thermalConductivity}
\end{equation}
\begin{equation}
{c_{v}}_{i} = {c_{p}}_{i}-R_{i}
\label{eq:specificHeat}
\end{equation}

\subsection{Electromagnetic multi-fluid}
A multi-fluid model \change[]{was}{is} used for the interaction of a radio wave with the plasma. Maxwell's equations \change[]{were}{are} used to solve for the evolution of the electric and magnetic fields. Ampere's law and Faraday's law are given by Eqs. (\ref{eq:ampere}) and (\ref{eq:faraday}) respectively. The right hand side of Eq. (\ref{eq:ampere}) is the sum of the current densities of the conducting \change[]{species}{fluids}. \add[]{The upper limit $N$ in the summation is the total number of fluids in the system.} The divergence equations (\ref{eq:electricDivergence}) and (\ref{eq:magneticDivergence}) should be satisfied along with the Ampere's and Faraday's laws.\cite{munz2000athreedimensional}

\begin{equation}
\frac{\partial \vec{E}}{\partial t}-c^{2}\nabla\times\vec{B} = -\frac{1}{\epsilon_{0}}\sum\limits_{\alpha}^{N}\frac{q_{\alpha}\rho_{\alpha}\vec{u}_{\alpha}}{m_{\alpha}}
\label{eq:ampere}
\end{equation}

\begin{equation}
\frac{\partial \vec{B}}{\partial t}+\nabla\times\vec{E} = 0
\label{eq:faraday}
\end{equation}

\begin{equation}
\nabla\cdot\vec{E}= \frac{1}{\epsilon_{0}}\sum\limits_{\alpha}^{N}\frac{q_{\alpha}\rho_{\alpha}}{m_{\alpha}}
\label{eq:electricDivergence}
\end{equation}

\begin{equation}
\nabla\cdot\vec{B}= 0
\label{eq:magneticDivergence}
\end{equation}

The transport of the multi-fluid system \change[]{was}{is} modeled using the system of equations Eqs. (\ref{eq:twoFluidMass})--(\ref{eq:twoFluidTotalEnergy}). Index $\alpha$ is for any fluid. The first \change[]{terms}{term} on the RHS of Eq.(\ref{eq:twoFluidMomentum}) \add[]{represents} the electric and magnetic Lorentz forces. The third term is the net momentum exchange with the remaining fluids in the system.\cite{zhdanov2002transport} The first term on the RHS of Eq.(\ref{eq:twoFluidTotalEnergy}) is for Joule heating and the fourth and fifth terms are kinetic energy and internal energy exchange terms respectively.\cite{zhdanov2002transport} \add[]{The bulk velocity $V$, momentum exchange term $R_{\alpha}$, and the internal energy exchange term $Q_{\alpha}$ are given by Eqs. } (\ref{eq:bulkVelocity})--(\ref{eq:intEnergyEx}) \add[]{respectively.}
  
\begin{equation}
\frac{\partial \rho_{\alpha}}{\partial t} + \nabla\cdot\left(\rho_{\alpha} \vec{u}_{\alpha}\right) = 0
\label{eq:twoFluidMass}
\end{equation}
\begin{equation}
\frac{\partial \left(\rho_{\alpha} \vec{u}_{\alpha}\right)}{\partial t} + \nabla\cdot\left(\rho_{\alpha} \vec{u}_{\alpha}\vec{u}_{\alpha} + p_{\alpha}I\right) = \frac{\rho_{\alpha}}{m_{\alpha}} q_{\alpha} \left(\vec{E}+\vec{u}_{\alpha}\times\vec{B} \right) + \nabla\cdot\mathbf{\tau}_{\alpha} + \vec{R}_{\alpha}
\label{eq:twoFluidMomentum}
\end{equation}
\begin{equation}
\frac{\partial \left(e_{\alpha}\right)}{\partial t} + \nabla\cdot\left(\vec{u}_{\alpha}\left(e+p_{\alpha}\right)\right) = \frac{\rho_{\alpha}}{m_{\alpha}} q_{\alpha}\vec{u}_{\alpha}\cdot\vec{E} + \nabla\cdot\left(\tau_{\alpha}\cdot\vec{u}_{\alpha}\right) + \nabla\cdot\left(k_{\alpha}\nabla T_{\alpha}\right) + \vec{V}_{\alpha}\cdot\vec{R}_{\alpha} + Q_{\alpha}
\label{eq:twoFluidTotalEnergy}
\end{equation}
where, 
\begin{equation}
\vec{V} = \left(\sum\limits_{i}^{N}\rho_{i}\vec{u}_{i}\right)/\sum\limits_{i}\rho_{i}
\label{eq:bulkVelocity}
\end{equation}
\begin{equation}
\vec{R}_{\alpha} = - \sum\limits_{i\ne\alpha}^{N}\frac{\rho_{\alpha}}{m_{\alpha}}\mu_{\alpha i}\zeta_{\alpha i}^{-1}\left(\vec{u}_{\alpha}-\vec{u}_{i}\right) 
\end{equation}
and
\begin{equation}
Q_{\alpha} = - \sum\limits_{i\ne\alpha}^{N}3k_{B}\frac{\rho_{\alpha}}{m_{\alpha}}\left[\mu_{\alpha i}/\left(m_{\alpha}+m_{i}\right)\right]\zeta_{\alpha i}^{-1}\left(T_{\alpha}-T_{i}\right)
\label{eq:intEnergyEx}
\end{equation}

\change[]{i}{I}mportantly, this model describes electromagnetic wave propagation in free space (\change[]{when}{where} the charged species densities are zero) and in a conducting fluid (\change[]{when}{where} the charged species densities are non-zero).  In particular it describes reflection of electromagnetic waves off of an over-dense plasma as well as electromagnetic wave propagation in a plasma including the changes caused by external magnetic fields.  The model is more complete to that of magnetohydrodynamics (MHD) and can be thought of as MHD without the assumption of quasi-neutrality, without the assumption that the light wave is infinitely fast and by using the full Ohm's law, (including electron inertia) in the MHD system.  Restricting ourselves to two-fluids for the moment (electrons and ions only), 4 parameters that can be derived from this model, will be important in determining the electromagnetic wave propagation characteristics in the plasma.  The first is the electron plasma frequency
\begin{equation}
\Omega_{p\,e}=\sqrt{\frac{n_{e}q_{e}^{2}}{m_{e}\epsilon_{0}}}
\end{equation}
the second is the electron cyclotron frequency
\begin{equation}
\Omega_{c\,e}=\frac{q_{e}\,B_{0}}{m_{e}}
\end{equation}
followed by the ion plasma frequency
\begin{equation}
\Omega_{p\,i}=\sqrt{\frac{n_{i}q_{i}^{2}}{m_{i}\epsilon_{0}}}
\end{equation}
and the ion cyclotron frequency
\begin{equation}
\Omega_{c\,i}=\frac{q_{i}\,B_{0}}{m_{i}}
\end{equation}
These parameters will be used in the discussion of the whistler wave.

\add[]{The system of Eqs. }(\ref{eq:mass})--(\ref{eq:specificHeat}) \add[]{will be used for the simulation of the bulk fluid and species transport on the RAM C, which will be discussed in Sec. }\ref{sec:reactiveFlow}\add[]{. The species density obtained will serve as input to the blackout analysis. The em wave propagation, reflection by the plasma layer, and whistler mode propagation will be simulated using the Eqs. }(\ref{eq:ampere})--(\ref{eq:intEnergyEx}) \add[]{in Secs. }\ref{sec:dispersion} and \ref{sec:emPropagation}\add[]{. It has to be noted here that, only a simplified version of Eqs. }(\ref{eq:twoFluidMass})--(\ref{eq:twoFluidTotalEnergy})\add[]{ are used for the fluids transport}. \add[]{The plasma frequency time scale is much smaller than the advection, diffusion and collision time scales, and hence these terms can be neglected. A detailed discussion on the frequencies is given in Sec. } \ref{sec:emPropagation}.

\subsection{Solution methodology}

The equation systems given in Sec. \ref{sec:mathematicalFormulation} \change[]{were}{are} solved using a generalized unstructured grid finite volume solver, USim\cite{{loverich2013nautilus},{shashurin2014laboratory},{kundrapu2013modeling}}.  Though multi-fluid electromagnetic solvers have been developed throughout the years by several researchers \cite{{shumlak2003approximate},{hakim2006high},{loverich2011},{kumar2012entropy},{thompson2013fully},{johnson2013gaussian},{shumlak2013high},{srinivasan2011numerical}}, the present solver is the first solver using an unstructured formulation and running on an unstructured grid\cite{loverich2013nautilus} as prior codes were based on multi-block logically Cartesian grids.  The flux reconstruction on the cell faces \change[]{was}{is} carried out using second order accurate Monotonic Upstream-Centered Scheme for Conservation Laws (MUSCL)\cite{vanLeer1979towards}. The right and left fluxes on the cell faces were obtained by extrapolating the cell centered gradient of the conserved variable. The spurious oscillations that may arise due to the flux reconstruction are limited using flux limiters such as Van-Leer limiter.\cite{vanLeer1979towards} The cell centered gradient \change[]{was}{is} computed using the weighted least squares method. A second degree polynomial \change[]{was}{is} considered in this work.  The actual flux on the cell face \change[]{was}{is} then obtained using an approximate Riemann flux.  In this work HLLE approximate Riemann flux\cite{einfeldt1988godunov} \change[]{was}{is} considered for the fluid equations while full wave flux \change[]{was}{is} chosen for the Maxwell's equations. The diffusion fluxes \change[]{were}{are} evaluated by computing the least squares gradient on the cell faces and then performing a surface integral according to the Gauss divergence theorem. \add[]{The bulk fluid velocity is used to advect the species in in Eq. }(\ref{eq:speciesCont})\add[]{. The same flux scheme is used as that of the bulk fluid.} The \change[]{reaction rates were}{species density sources due to chemical reactions are} integrated separately using the Boost ODE integrator\cite{ahnert2011odeint}. The left and right hand side of the Eqs. (\ref{eq:mass})--(\ref{eq:speciesCont}), (\ref{eq:ampere}), and (\ref{eq:faraday}) \change[]{were}{are} evaluated separately and added together and then integrated in time using Runge-Kutta method. A third order RK integration scheme \change[]{was}{is} used.  Operator splitting \change[]{was}{is} used for the reaction terms and super time stepping for the diffusion terms.
 
\section{Validation of the results}
\subsection{Reactive flow simulation}
\label{sec:reactiveFlow}
\begin{figure}
\begin{center}
\includegraphics[width=.4\textwidth]{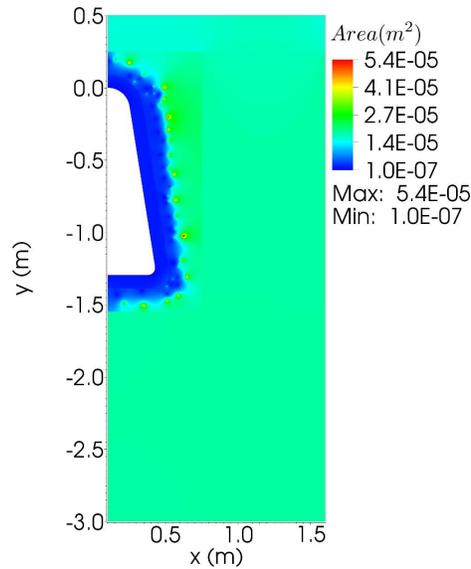}
\caption[]{The unstructured grid used for the simulation. The contour flood represents the cell area.}
\vspace{-0.2in}
\label{fig:grid}
\end{center}
\end{figure}

The simulation \change[]{was}{is} performed for a velocity of 7650 $m/s$, density  2.816$\times10^{-4}$ $kg/m^{3}$ and temperature 244.3 K. These values correspond to an altitude of 61 $km$. The air species $N_2$, $O_2$, $N$, $O$, $NO$, $NO^{+}$, and $e$ (electron) \change[]{were}{are} considered for the reaction chemistry and the reaction rates \change[]{were}{are} considered from the Ref. \cite{josyula2003governing}. The unstructured grid used for the simulation \change[]{was}{is} shown in Fig. (\ref{fig:grid}). \add[]{The geometry of RAM C vehicle is available in Ref.} \cite{jones1972electrostatic}. \add[]{RAM C is a blunt cone ballistic body with a nose cap radius of 15.24 cm, half angle} $9^{\circ}$\add[]{, and a length of 1.29 m. The probes mounted on the surface are not modeled in this work.} Cubit \cite{blacker1994cubit} grid generation software \change[]{was}{is} used for the grid generation. The contour flood represents the area of cells in $m^{2}$. The average edge lengths vary from around 0.5 $mm$ at the nose cap region to 1 $mm$ on the lateral surface.

\begin{figure}
\begin{center}
\subfigure[]{
\includegraphics[width=.33\textwidth]{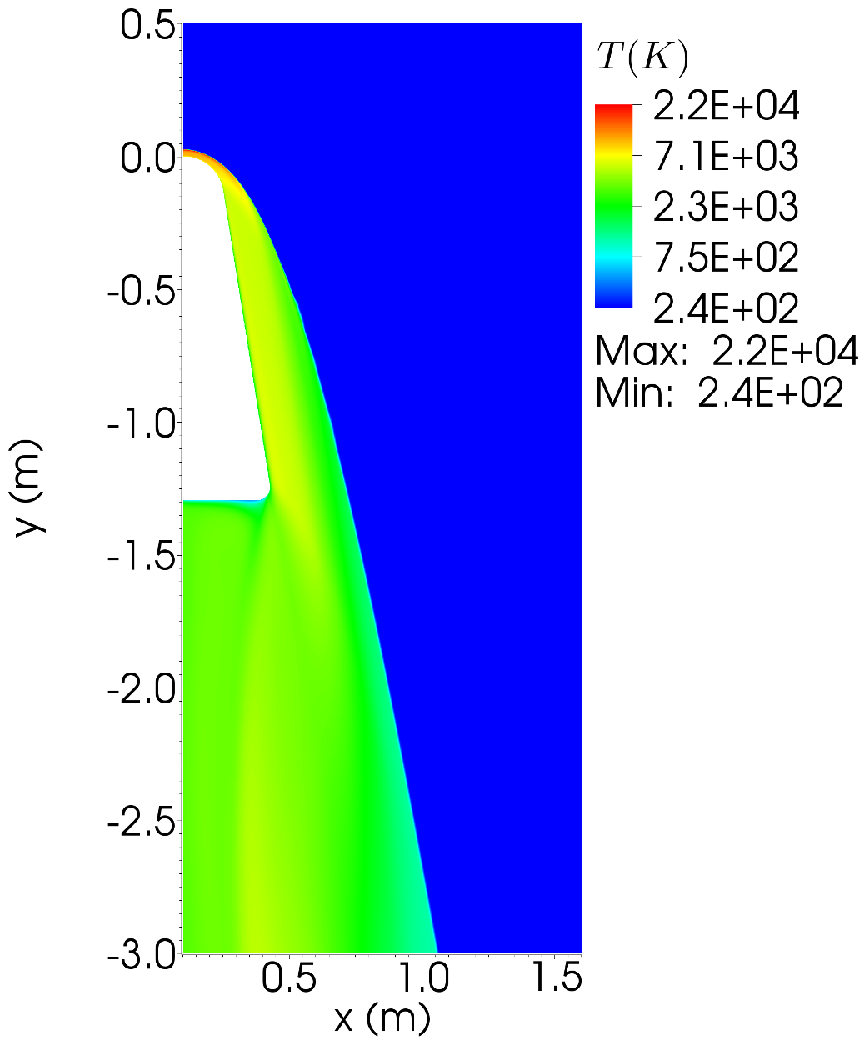}
\label{fig:ramcTemp}
}
\quad
\subfigure[]{
\includegraphics[width=.33\textwidth]{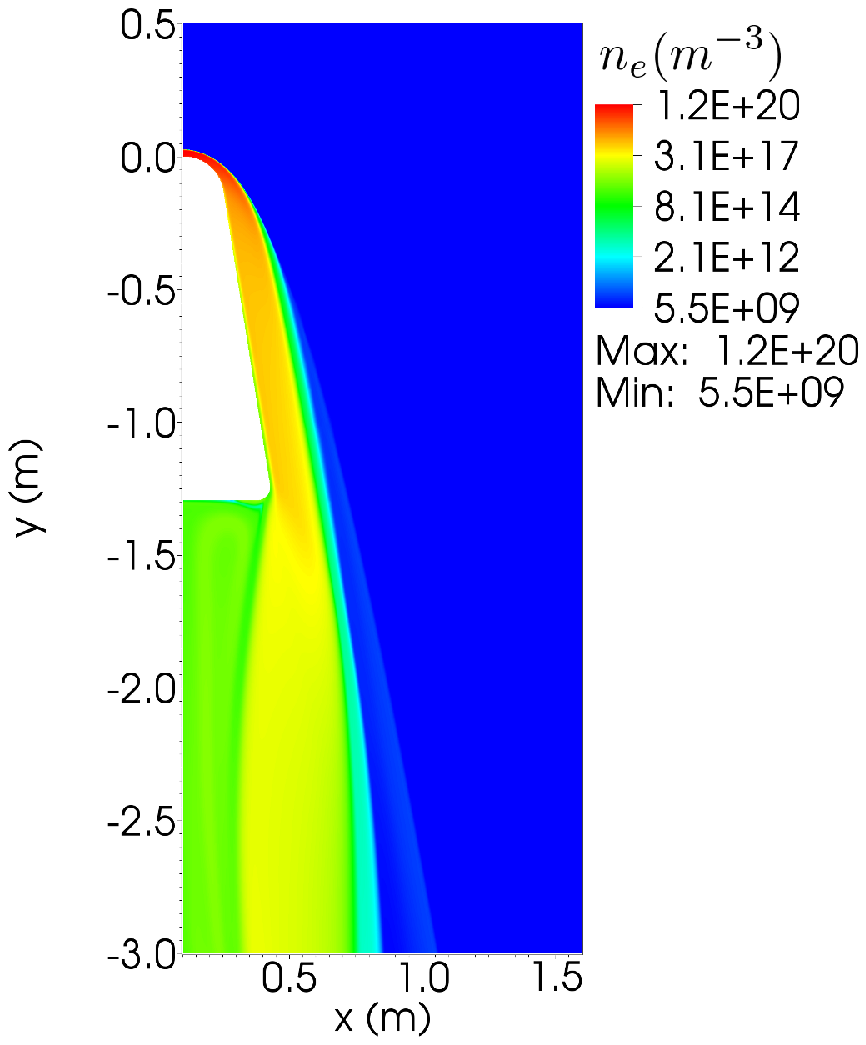}
\label{fig:ramcNe}
}
\caption[]{Flow parameters on RAM C. \subref{fig:ramcTemp} Temperature distribution and \subref{fig:ramcNe} electron number density distribution.}
\vspace{-0.2in}
\label{fig:ramcParameters}
\end{center}
\end{figure}

\begin{figure}
\begin{center}
\includegraphics[width=.5\textwidth]{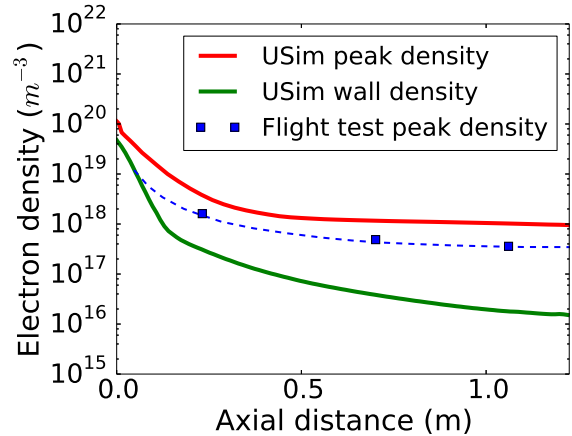}
\caption[]{Validation of the electron density in the plasma layer of RAM C. The top and bottom curves are peak density and surface density respectively. The middle curve is the averaged peak density measurements from Ref.\cite{jones1972electrostatic}}
\vspace{-0.2in}
\label{fig:peakDensCompare}
\end{center}
\end{figure}

Flow enters the domain from the top boundary of Fig. \ref{fig:grid}.  An axisymmetric boundary condition \change[]{was}{is} imposed on the axis (left boundary). On that wall, a standard no slip and radiation equilibrium temperature \change[]{were}{are} imposed. Outflow boundary conditions \change[]{were}{are} used on the remaining boundaries. The temperature and electron density distributions are shown in Fig. \ref{fig:ramcParameters}. \add[]{The species are transported with the same advection velocity of bulk fluid. The species densities at the top boundary are equal to their free stream densities and the gradient of species density is zero on the remaining boundaries.} The peak values of average temperature and the electron density are 21860 K and $1.18\times10^{20}$ m$^{-3}$ respectively existing in the stagnation region. Figure \ref{fig:peakDensCompare} shows the comparison of the surface and peak electron densities in the plasma layer with the reflectometer measurements presented in Ref. \cite{jones1972electrostatic}. The measurements represent the time averaged values of the electron density measured using 15 reflectometers of four different frequencies placed at four stations on the wall of RAM C. The cut-off densities associated with the four frequencies are $1.52\times10^{19}$,$1.25\times10^{18}$,$1.37\times10^{17}$, and $1.54\times10^{16}$ m$^{-3}$ respectively.\cite{grantham1970flight} The first station \change[]{was}{is} located at 0.0457 m from the nosecap tip. The remaining three stations locations along with the measured peak densities are shown by squares in the figure. The dashed curve represents the curve fit of the measured data points. The bottom most curve is the surface density distribution while the top most curve is the peak density of electrons in the plasma layer of Fig.\ref{fig:ramcNe}.  The peak values of the simulation are about three times the values from the measurements. Inclusion of radiation losses from the plasma and the diffusion of electrons in the simulation could decrease the density to some extent. Moreover, the comparison shows a good agreement in terms of the trend and in the design point view, the higher values of the simulation to this extent are acceptable in terms of the factor of safety. The wall densities are well below the peak values\remove[]{,} for the simple reason that the wall temperature is much below the boundary layer temperature. 

\subsection{Dispersion relation for waves parallel to the magnetic field}
\label{sec:dispersion}
The magnetic window approach to radio communication blackout mitigation takes advantages of special properties of electromagnetic wave propagation in plasmas in the presence of a magnetic field.  A derivation of the whistler wave (as well as many other plasma waves) can be found in many plasma physics text books including \cite{goldston2000introduction}.  In the case below, the dispersion relation is derived from the two-fluid electromagnetic plasma model and written in terms of plasma parameters, $\Omega_{c\,e}$,$\Omega_{c\,i}$,$\Omega_{p\,e}$,$\Omega_{p\,i}$ and the speed of light $c$, the results are given as follows.  

The R-Mode dispersion relation is given by
\begin{equation}
k^{2}=\frac{w^2 \left(w^2+w (\Omega
   _{ci}-\Omega_{ce})-\Omega_{ce} \Omega_{ci}- \Omega_{pe}^2-
   \Omega_{pi}^2\right)}{c^2 (w-\Omega_{ce})
   (w+\Omega_{ci})}
\end{equation}
after ignoring the ion motion,  the R-Mode dispersion relation becomes
\begin{equation}
k^{2}=\frac{w \left(w^2 -w
   \Omega_{ce}-
   \Omega_{pe}^2\right)}{c^2 (w-\Omega_{ce})}
   \label{eq:rmode}
\end{equation}.

\begin{figure}
\begin{center}
\includegraphics[width=.45\textwidth]{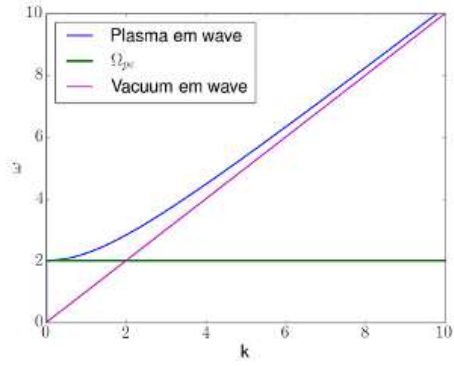}
\caption[]{Dispersion graph for electromagnetic waves traveling in a plasma with no background magnetic field, frequency vs wave number in normalized units.}
\vspace{-0.2in}
\label{fig:dispersionB0}
\end{center}
\end{figure}

\begin{figure}
\begin{center}
\includegraphics[width=.45\textwidth]{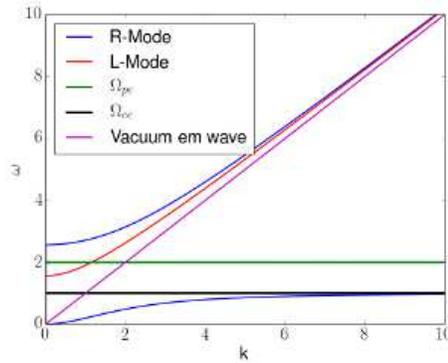}
\caption[]{Dispersion graph for electromagnetic waves traveling parallel to the magnetic field, frequency vs wave number in normalized units.}
\vspace{-0.2in}
\label{fig:dispersion}
\end{center}
\end{figure}

The L-Mode dispersion relation is given by
\begin{equation}
k^{2}=\frac{w^2 \left(w^2 +w
    (\Omega_{ce}-\Omega_{ci})-\Omega_{ce}
   \Omega_{ci}-
   \Omega_{pe}^2-
   \Omega_{pi}^2\right)}{c^2 (w+\Omega_{ce}) (w-\Omega_{ci})}
\end{equation}.
after ignoring the ion motion,  the L-Mode dispersion relation becomes
\begin{equation}
k^{2}=\frac{w \left(w^2 +w
   \Omega_{ce}-\Omega_{ce}
   \Omega_{pe}^2\right)}{c^2 (w+\Omega_{ce})}
\end{equation}
Figure \ref{fig:dispersionB0} shows the frequency vs wave number plotted for the electromagnetic wave in a plasma without a background magnetic field.  The vacuum electromagnetic wave is provided for reference.  The plasma electromagnetic wave does not propagate below the plasma frequency represented by the horizontal line. This is the reason for radio communication blackout. At all frequencies below the plasma frequency the wave is evanescent.  By adding a magnetic field the dispersion relation changes and the wave can propagate through the plasma at frequencies below the electron cyclotron frequency.
Figure \ref{fig:dispersion} shows the frequency vs wave number plotted for the R-Mode and L-Mode waves in non-dimensional units in a plasma with a background magnetic field. The R-mode wave has two branches, the lower frequency branch which propagates below the plasma frequency is known as the whistler wave.  The whistler wave has a cutoff at the electron cyclotron frequency.  In the presence of a magnetic field then it is possible for the electromagnetic wave to penetrate the over-dense plasma.  The electron cyclotron frequency must be greater than the signal frequency and this puts a lower bound on the magnetic field strength that should be used. Based on the electromagnetic wave frequency, the background field can be adjusted so that the signal can propagate through the plasma as a whistler wave.  This method is termed as the magnetic window in the blackout mitigation studies.

\begin{figure}
\begin{center}
\subfigure[]{
\includegraphics[width=0.45\textwidth]{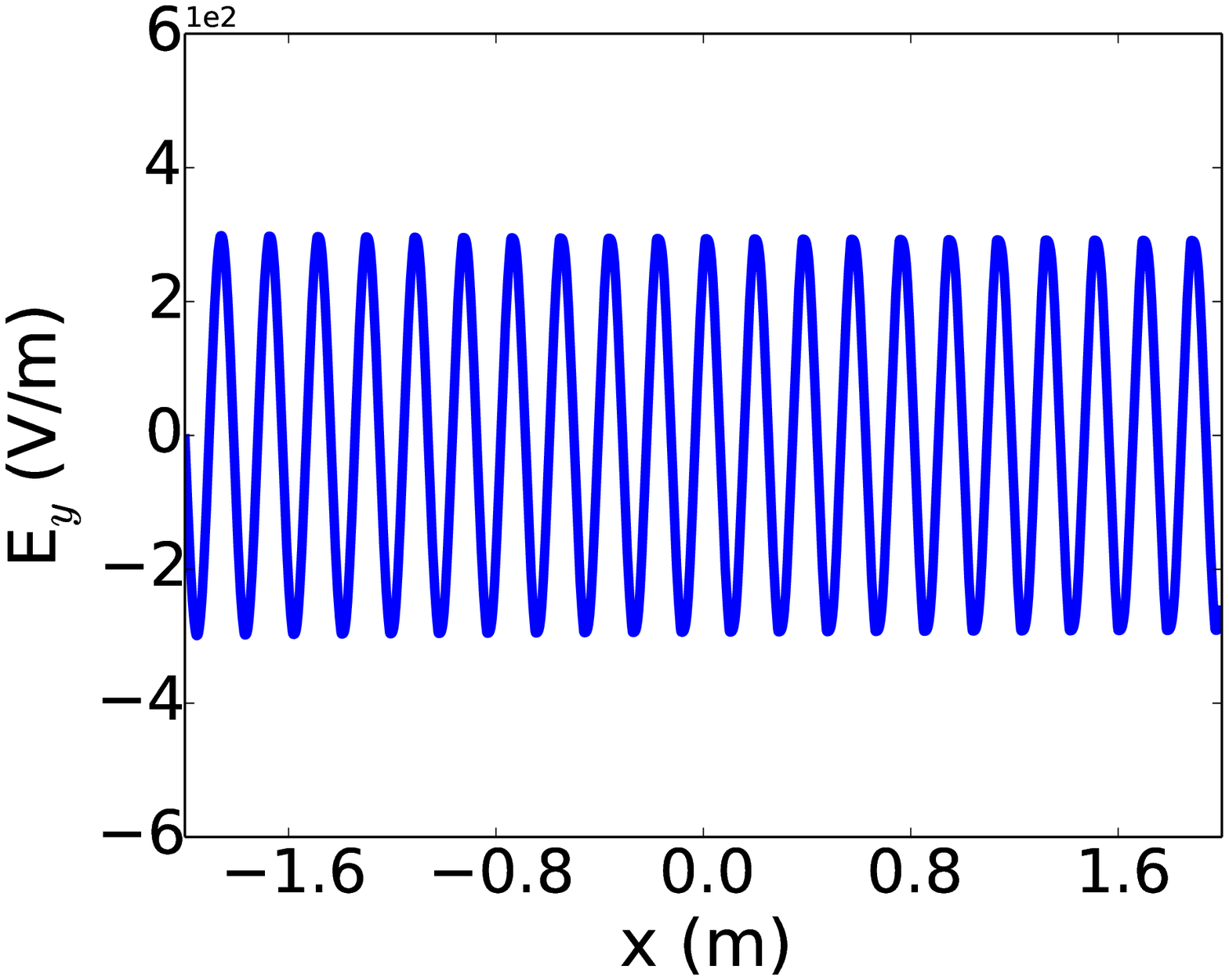}
\label{fig:eyVacuum}}
\quad
\subfigure[]{
\includegraphics[width=0.45\textwidth]{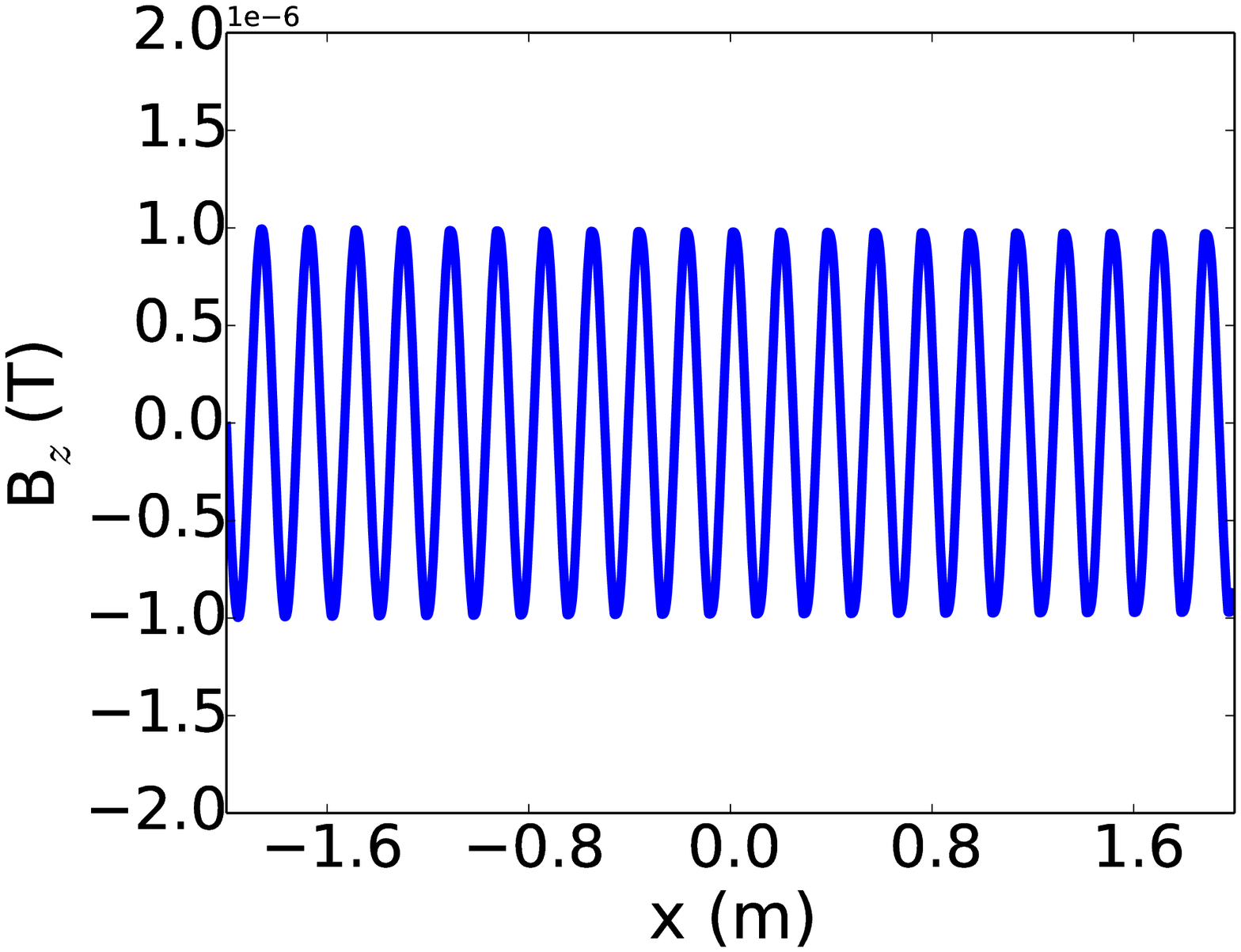}
\label{fig:bzVacuum}}
\caption[]{The \subref{fig:eyVacuum} y component of electric field and the \subref{fig:bzVacuum} z component of the magnetic field in the EM wave of frequency 1.6 GHz traveling in neutral fluid. }
\vspace{-0.2in}
\label{fig:emVacuum}
\end{center}
\end{figure}

\begin{figure}
\begin{center}
\subfigure[]{
\includegraphics[width=.45\textwidth]{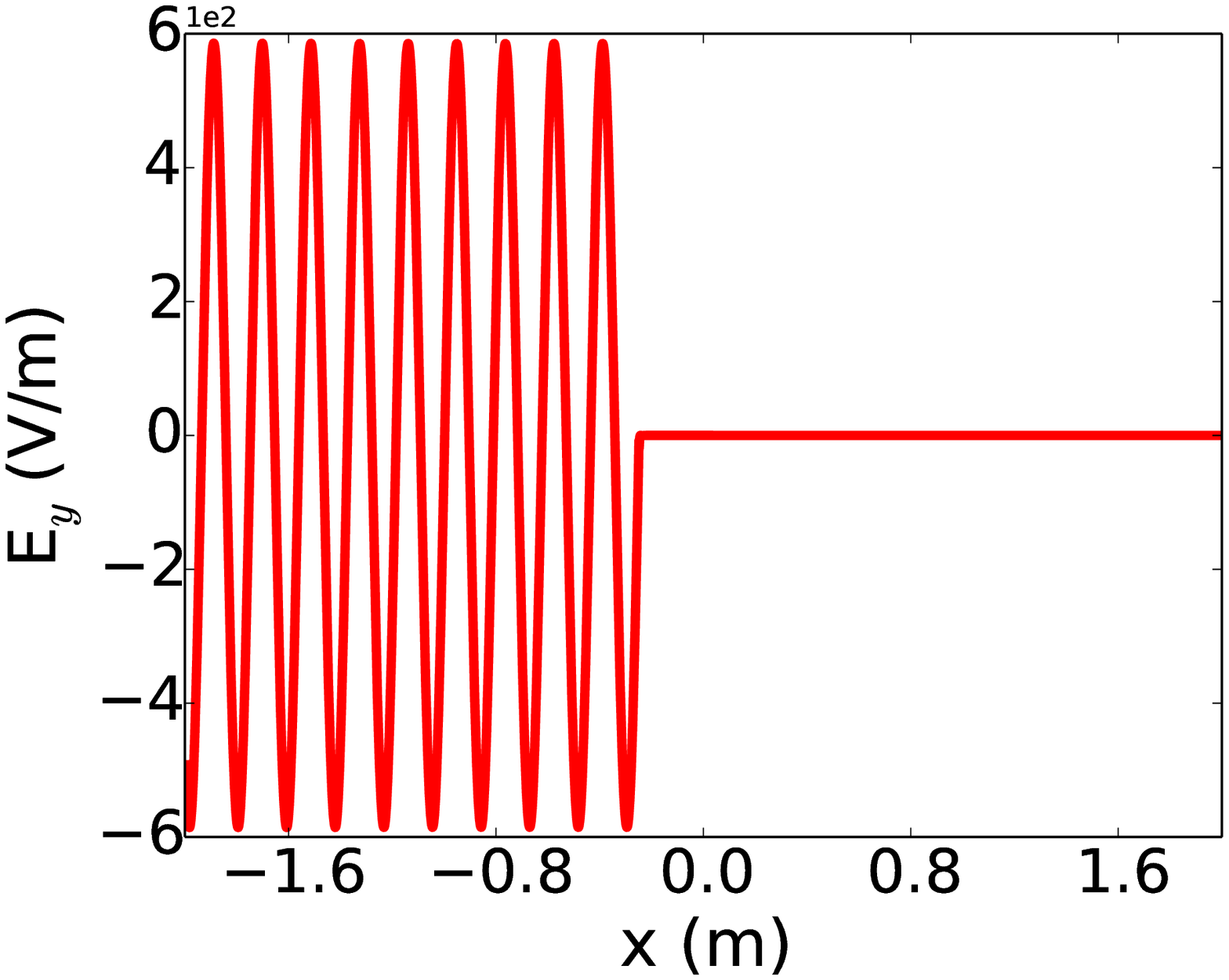}
\label{fig:eyB0TNe1e19}}
\quad
\subfigure[]{
\includegraphics[width=.45\textwidth]{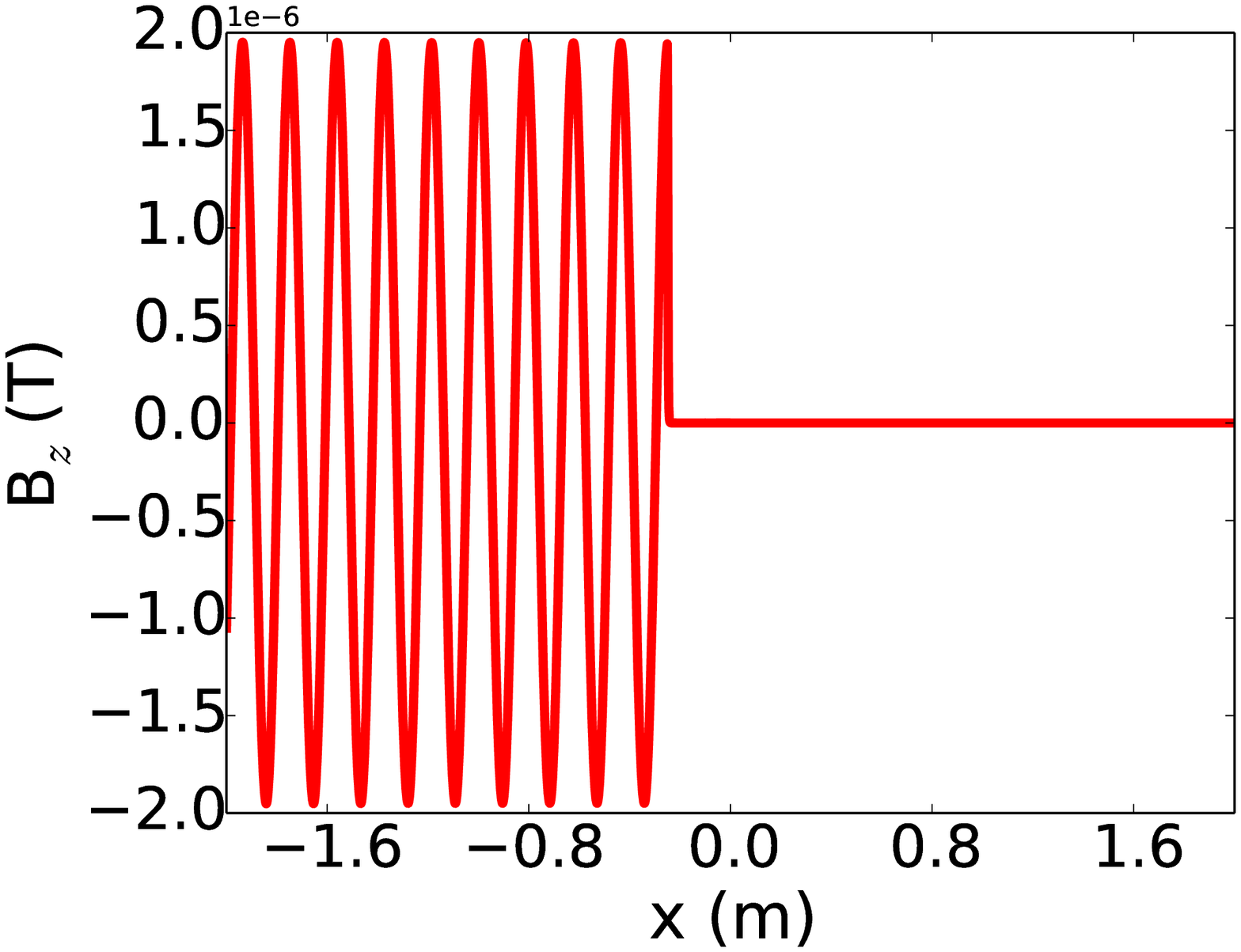}
\label{fig:bzB0TNe1e19}}
\caption[]{The \subref{fig:eyB0TNe1e19}y component of electric field and the \subref{fig:bzB0TNe1e19} z component of the magnetic field of the EM wave of frequency 1.6 GHz traveling in a domain with plasma slab of thickness 0.3 m. The slab starts from x = -0.15 m.}
\vspace{-0.2in}
\label{fig:emB0TNe1e19}
\end{center}
\end{figure}

\add[]{The electromagnetic multi-fluid Eqs. }(\ref{eq:ampere})--(\ref{eq:magneticDivergence}) and (\ref{eq:twoFluidMomentum})--(\ref{eq:twoFluidTotalEnergy}) \add[]{are solved in one-dimension to propagate an EM wave in a neutral fluid and a plasma slab. The advection, diffusion and collision terms are neglected for ion and electron fluids. The simulation results are compared to the analytical calculations of the dispersion relation Eq. }(\ref{eq:rmode}). Figure \ref{fig:emVacuum} shows the EM wave propagation in a neutral fluid. A plane wave \change[]{was}{is} excited from the left boundary with the components E$_y$ = $c\,a_{0}sin(2\pi f t)$ and B$_z$ = E$_y$/c. The frequency of the wave \change[]{was}{is} 1.6 GHz. Uninterrupted propagation of the wave can be clearly seen. A uniform plasma slab of thickness 0.3 m was then added in the domain at x = -0.15 m. The plasma density \change[]{was}{is} $10^{19}$ m$^{-3}$. Since the frequency of the plasma, 28.4 GHz is much greater than the wave frequency, the wave \change[]{was}{is} completely reflected by the plasma. Figure \ref{fig:emB0TNe1e19} shows the reflection of electric and magnetic fields. \add[]{The reflected wave components} E$_{y}$ and B$_{z}$ are shown in Fig. \ref{fig:eyB0TNe1e19} and \ref{fig:bzB0TNe1e19} respectively. The amplitudes of the electric and magnetic fields doubled since a standing wave is formed upon reflection from the plasma slab. The figure also shows the simulation accuracy of the two-fluid solver.\cite{{jenkins2013time},{merritt2013experimental}}

\begin{figure}
\begin{center}
\subfigure[]{
\includegraphics[width=.45\textwidth]{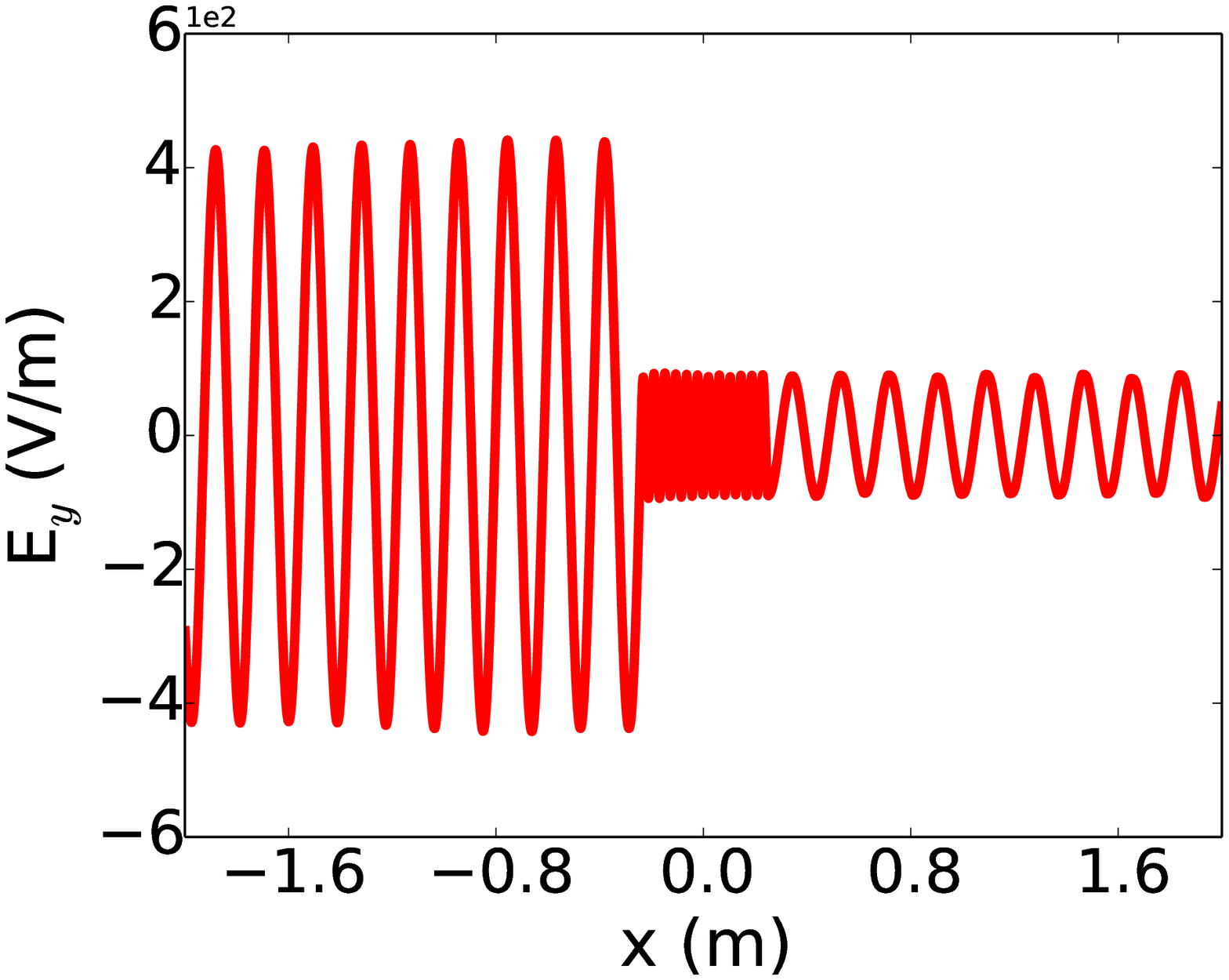}
\label{fig:ey1TPlasma}}
\quad
\subfigure[]{
\includegraphics[width=.45\textwidth]{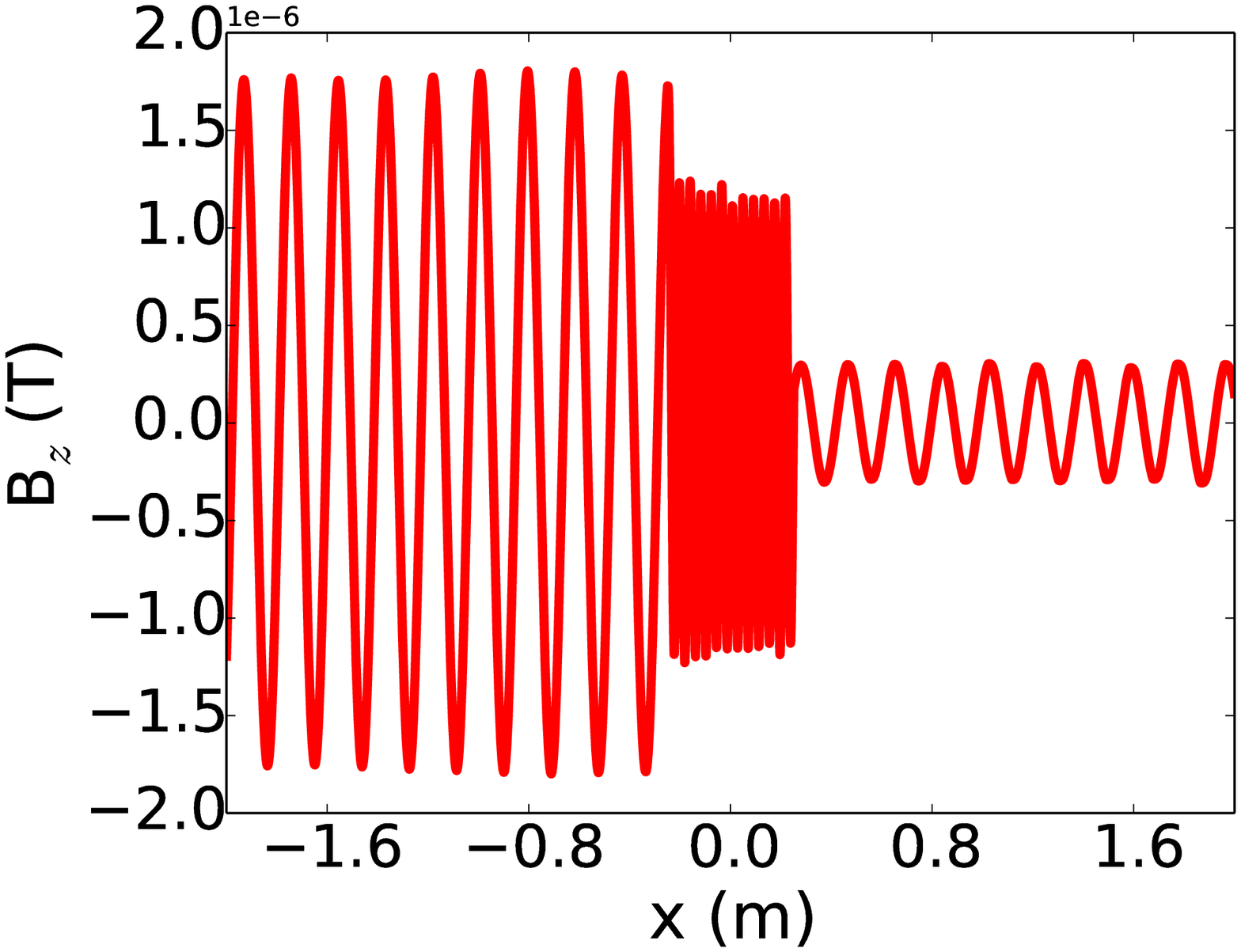}
\label{fig:bz1TPlasma}}

\caption[]{The \subref{fig:ey1TPlasma}y component of electric field and the \subref{fig:bz1TPlasma} z component of the magnetic field of the EM wave of frequency 1.6 GHz traveling in a domain with plasma slab of thickness 0.3 m. The slab starts from x = -0.15 m. A background magnetic field of B$_{0x}$ = 1 T is applied in the domain.}
\vspace{-0.2in}
\label{fig:eyB1TNe1e19}
\end{center}
\end{figure}

\begin{figure}
\begin{center}
\includegraphics[width=.5\textwidth]{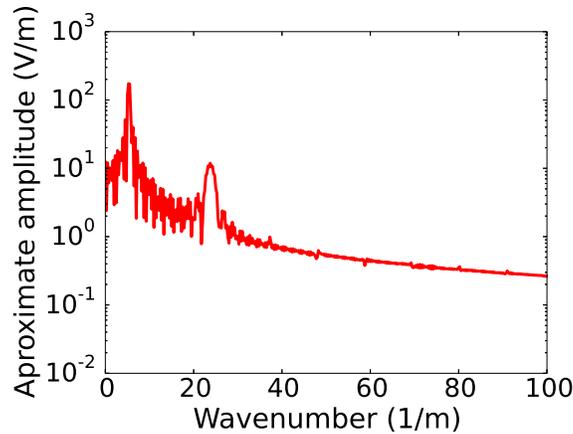}
\caption[]{The wave number of the plane wave in neutral fluid and plasma.}
\vspace{-0.2in}
\label{fig:waveNumber}
\end{center}
\end{figure}

A constant magnetic field of B$_{0x}$ \add[]{=} 1 T \change[]{was}{is} applied in the domain to create a magnetic window for the propagation of the wave in whistler mode through the plasma. \remove[]{Figure }\ref{fig:eyB1TNe1e19}\remove[]{ shows the wave components E$_y$} \add[]{Figure }\ref{fig:eyB1TNe1e19} \add[]{shows the the whistler wave propagation in the plasma slab.} \add[]{Figures} \ref{fig:ey1TPlasma} \add[]{and} \ref{fig:bz1TPlasma}  \add[]{show the wave components E$_y$ and B$_z$ respectively.} The whistler wave's accuracy is compared with the analytical solution obtained from the dispersion theory. The Dispersion relation shows that the wave number of the whistler wave in the plasma slab is 23.89. The Fourier transform of the wave in the spatial domain gives the wave number spectrum. \remove[]{The Fourier transform E$_x$ in Fig.}\ref{fig:eyB1TNe1e19}\remove[]{ is shown in Fig.}\ref{fig:waveNumber}\remove[]{.} \add[]{Figure }\ref{fig:waveNumber}\add[]{ shows the Fourier transform of the wave component E$_y$ of Fig.}\ref{fig:ey1TPlasma}. The first peak is located around k = 5.33 and  the second peak around k=23.89. The first peak represents the wave in the neutral zone and the second peak corresponds to the wave in the uniform plasma.  Note that the amplitude does not match with the values shown in Fig.\ref{fig:eyB1TNe1e19} since the exact wavenumbers 5.33 and 23.89 \change[]{were}{are} not resolved by the grid. \add[]{The current grid represents wave numbers in the multiples of 0.25. The exact peaks are located at 5.5 and 23.75. } A more refined grid \change[]{is required to represent}{that represents} the wave numbers of interest\change[]{, in which case, the}{would give a matching} spectral amplitude \change[]{will match the amplitude}{ to that} of the waves in the simulation. \add[]{Simulations are not performed on a further refined grid, since the wave number peaks are predicted with sufficient accuracy.} 
 
\section{Electromagnetic wave propagation over the RAM C}
\label{sec:emPropagation}
\subsection{Blackout}

\begin{figure}
\begin{center}
\subfigure[]{
\includegraphics[width=.4\textwidth]{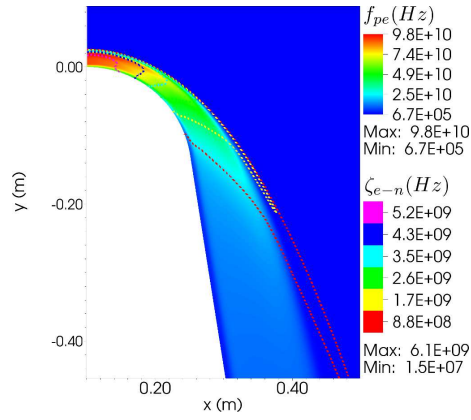}
\label{fig:collisionFrequencyContour}}

\quad

\subfigure[]{
\includegraphics[width=.4\textwidth]{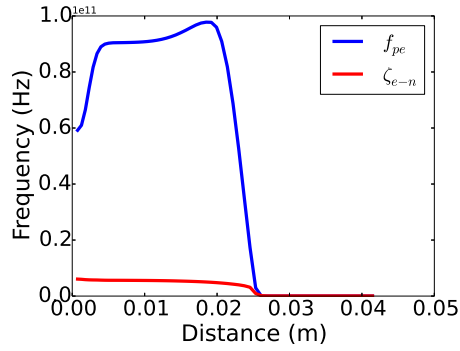}
\label{fig:collisionFrequencyLinePlot}}

\caption[]{Comparison of the plasma electron frequency \change[]{$\omega_{pe}$}{$f_{pe}$} and the collision frequency of electrons with neutrals $\zeta_{e-n}$. \subref{fig:collisionFrequencyContour} contour lines of $\zeta_{e-n}$ and the contour flood of \change[]{$\omega_{pe}$}{$f_{pe}$} and the \subref{fig:collisionFrequencyLinePlot} line plot of the frequencies along the stagnation line.}
\vspace{-0.2in}
\label{fig:frequencyTerms}
\end{center}
\end{figure}

For the current problem, in the Eqs.(\ref{eq:twoFluidMass})--(\ref{eq:twoFluidTotalEnergy}), the advection and viscous diffusion occur on much larger time scales when compared to the time scale of the plasma oscillations and EM wave. For instance, the smallest advection and diffusion time scales in the stagnation region are $4.8\times 10^{-6}$ and $8.9\times 10^{-6}$ s respectively. Whereas the plasma oscillations occur on the time scale of $1.1\times 10^{-11}$ s. In the aft region (y=-1.25m), the minimum advection and diffusion times are $2.73\times 10^{-7}$ and $1.78\times 10^{-6}$ s respectively. The plasma frequency time scale is $1.01\times 10^{-10}$ s. Overall, the timescales of the advection and diffusion are more than three orders of magnitude the plasma oscillation time scale. Hence, the advection and the viscous diffusion terms \change[]{were}{are} neglected. Note that the length scale used in the estimation of time scales \change[]{was}{is} the average edge length of the local cell.  The collision term in Eqs.(\ref{eq:twoFluidMass})--(\ref{eq:twoFluidTotalEnergy}) is neglected too as the collision frequency\cite{{rambo1994interpolation},{rambo1995comparison}} of the electrons and neutrals is less than an order of magnitude the plasma frequency. Figure \ref{fig:frequencyTerms} shows the comparison. The contour flood in Fig.\ref{fig:collisionFrequencyContour}represents the plasma electron frequency and the contour lines show the electron neutral collision frequency. In the present simulation, the electron neutral collision frequency is highest among the collisions of the remaining species. The highest values of the frequencies are seen near the stagnation region, where the densities are high. The contours clearly show that plasma frequency is higher than the collision frequency everywhere within the plasma layer. Also, a line plot comparison along the stagnation line is shown in Fig.\ref{fig:collisionFrequencyLinePlot} to get a better picture of the comparison of magnitudes.

\add[]{The Eqs. }(\ref{eq:twoFluidMass})--(\ref{eq:twoFluidTotalEnergy}) \add[]{(advection, diffusion and collision terms neglected) are solved for ion and electron transport. The boundary conditions for the ions and electrons are same as that of the bulk fluid mentioned in Sec.} \ref{sec:reactiveFlow}. \add[]{The electron and ion densities are initialized using the reactive flow solution. For the Maxwell's equations, conductor boundary condition is applied on the wall, which makes}  $\vec{n}\times\vec{E}$ \add[]{and} $\vec{n}\cdot\vec{B}$ \add[]{zero.}\cite{munz2000afinitevolume} \add[]{On all other boundaries, gradient of electric and magnetic fields is zero.}

\begin{figure}
\begin{center}
\subfigure[]{
\includegraphics[width=.3\textwidth]{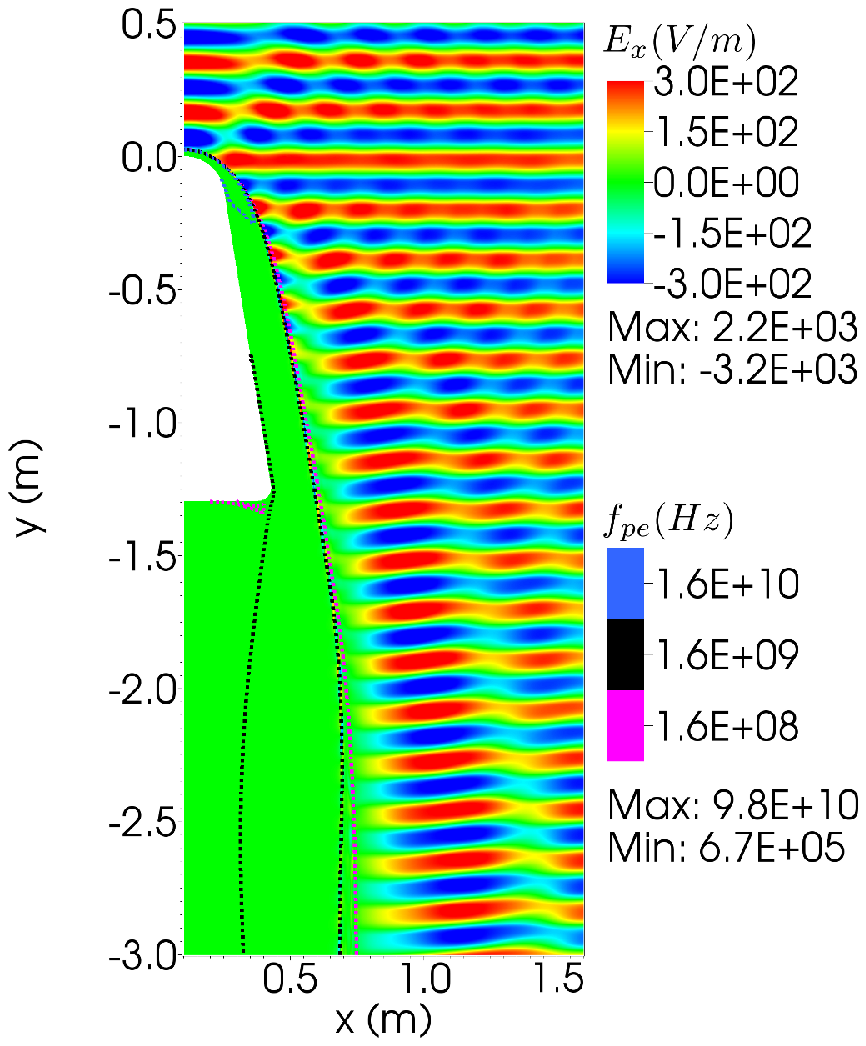}
\label{fig:Ex}}
\quad
\subfigure[]{
\includegraphics[width=.3\textwidth]{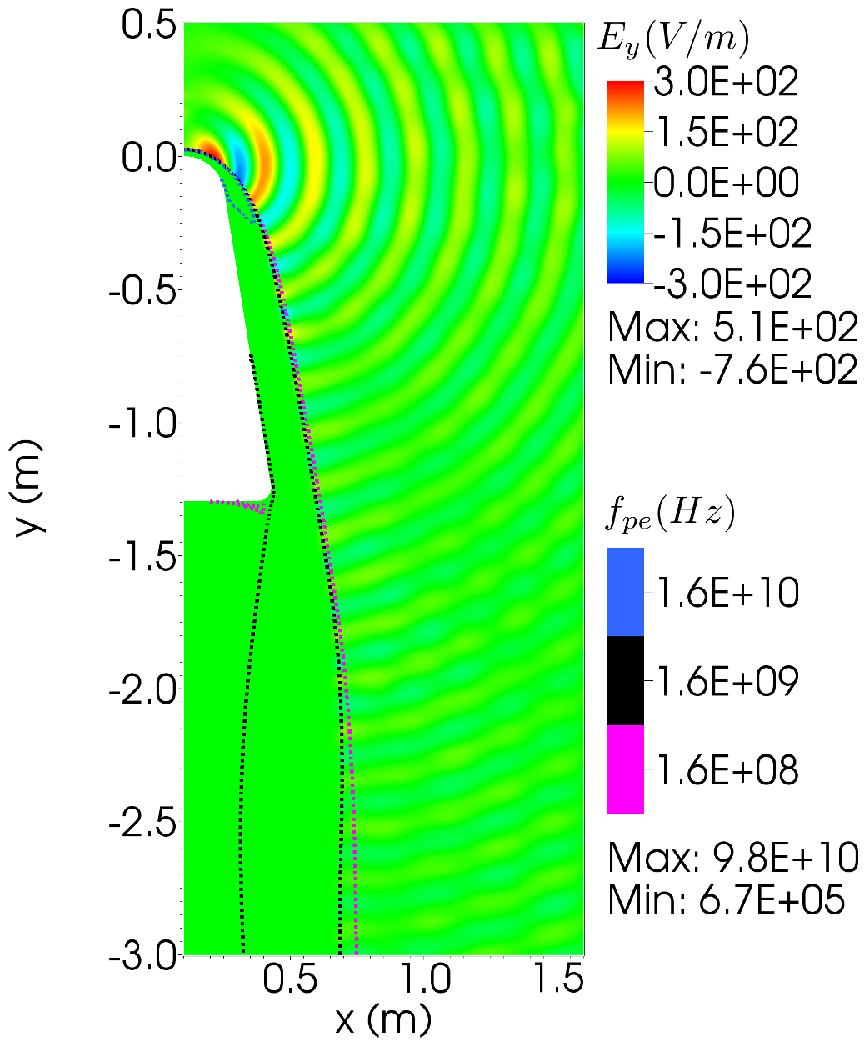}
\label{fig:Ey}}
\quad
\subfigure[]{
\includegraphics[width=.3\textwidth]{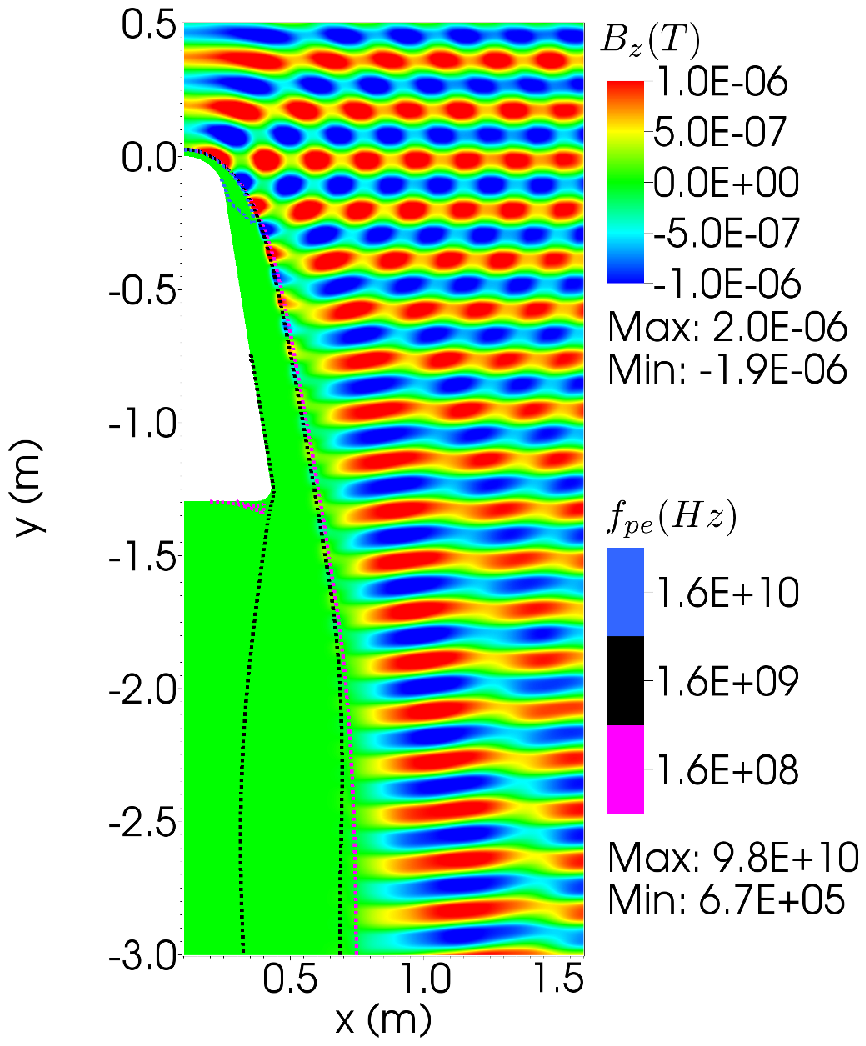}
\label{fig:Bz}}

\caption[]{EM wave reflection in the plasma layer of RAM C. \subref{fig:Ex} x component of the electric field, \subref{fig:Ey} y component of the electric field, and \subref{fig:Bz} z component of the magnetic field. \add[]{E$_z$, B$_x$, and B$_y$ are zero.} The contour lines represent the plasma electron frequency.}
\vspace{-0.2in}
\label{fig:blackout}
\end{center}
\end{figure}

The wave reflection by the plasma layer of the RAM C is shown in Fig.\ref{fig:blackout}. The frequency of the plane wave originating at the top boundary is 1.6 GHz. The wave components at the top boundary are  E$_x$ = $c a_{0}sin(2\pi f t)$, B$_z$ = E$_y$/c and the remaining components are equal to zero. Figures \ref{fig:Ex} and \ref{fig:Ey} represent the x,y components of the electric field. Figure \ref{fig:Bz} represents the z-component of the magnetic field. The contour flood shows the amplitudes of the wave. The dashed contour lines represent the plasma frequency. \add[]{The inner most contour line seen near the nosecap corresponds to 16 GHz. The middle and outer most contour lines correspond to 1.6 Ghz and 0.16 GHz respectively.} It can be clearly observed from the figures that the wave is completely reflected by the plasma layer once the plasma frequency is 1.6 GHz. Note that the flood contour levels are limited between peak positive and negative amplitudes of the original wave, in order to make the wave visible. The wave's amplitude increases by about 10 times at the edge of the plasma layer due to the resonance of the evanescent wave.\cite{white1974amplification} The amplified wave propagates along the plasma layer's edge.

\subsection{Magnetic window whistler mode}

\begin{figure}
\begin{center}
\subfigure[]{
\includegraphics[width=.3\textwidth]{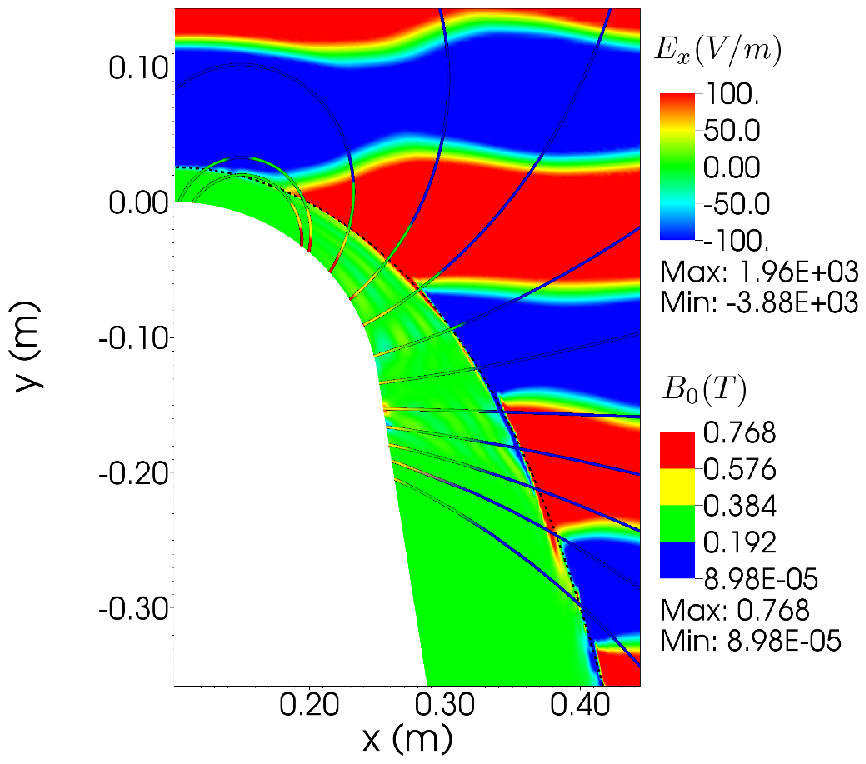}
\label{fig:lowWhistlerExNew}}
\quad
\subfigure[]{
\includegraphics[width=.3\textwidth]{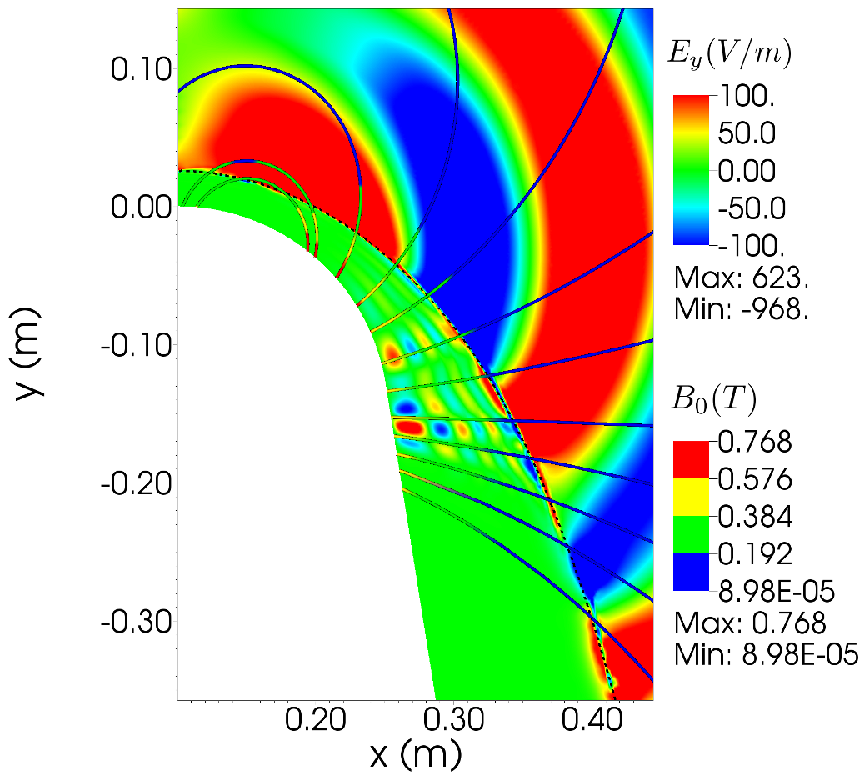}
\label{fig:lowWhistlerEyNew}}
\quad
\subfigure[]{
\includegraphics[width=.3\textwidth]{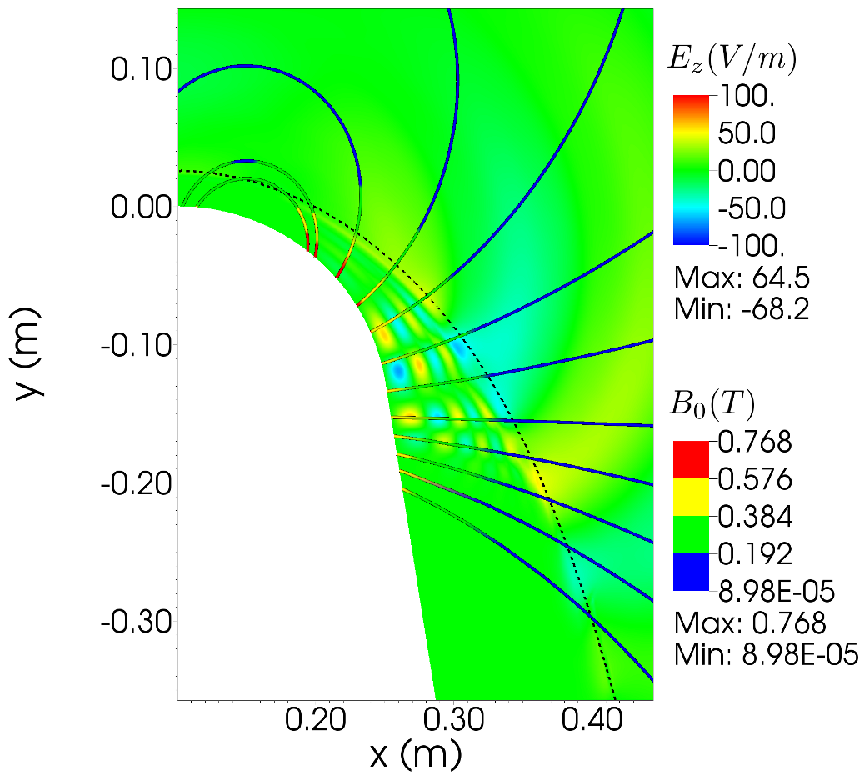}
\label{fig:lowWhistlerEzNew}}
\quad
\subfigure[]{
\includegraphics[width=.3\textwidth]{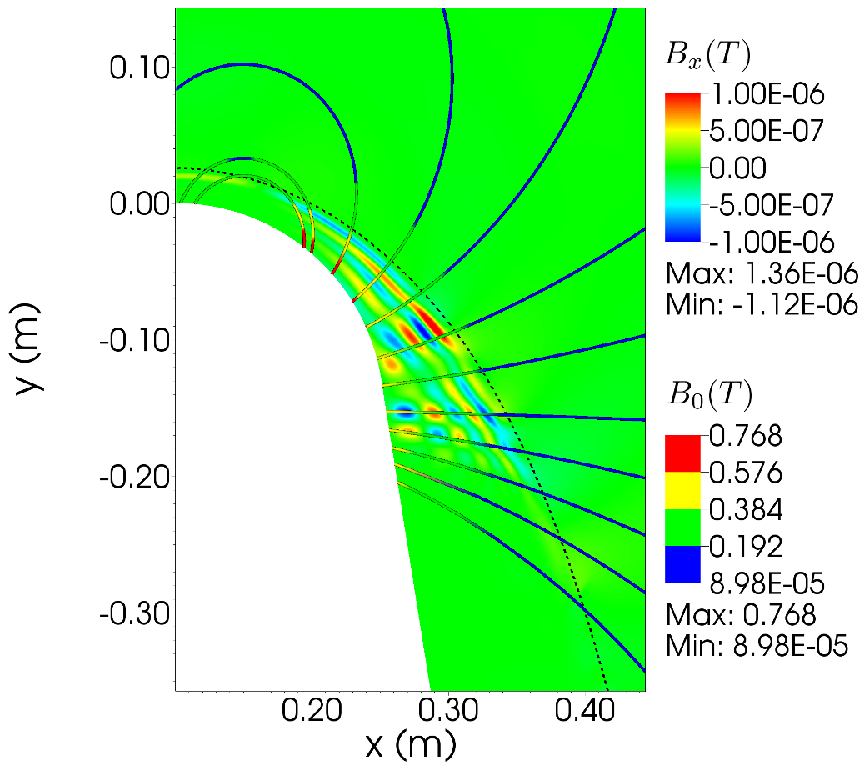}
\label{fig:lowWhistlerBxNew}}
\quad
\subfigure[]{
\includegraphics[width=.3\textwidth]{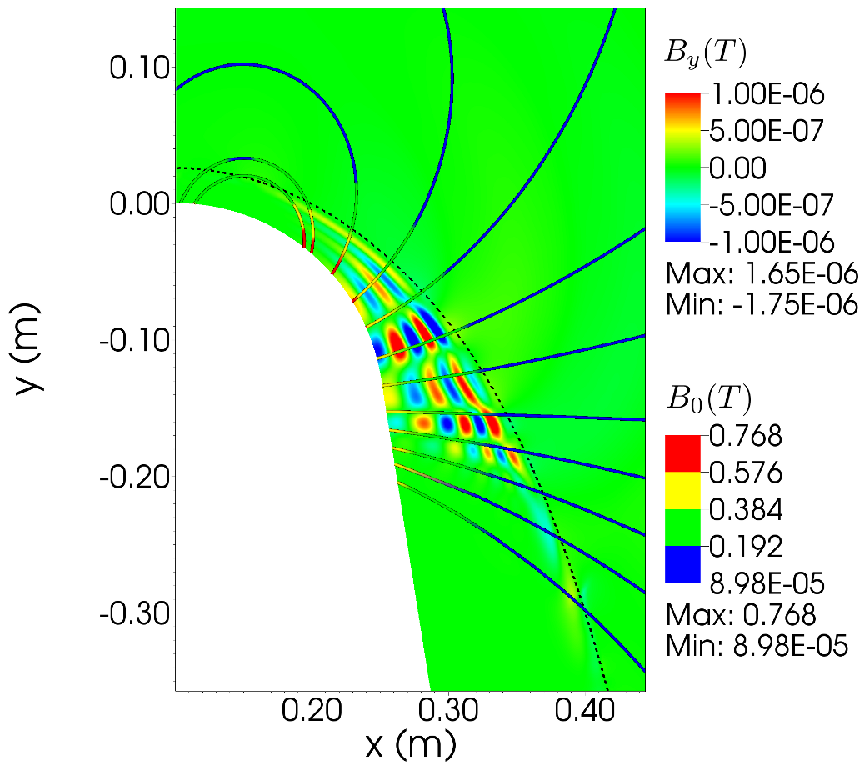}
\label{fig:lowWhistlerByNew}}
\quad
\subfigure[]{
\includegraphics[width=.3\textwidth]{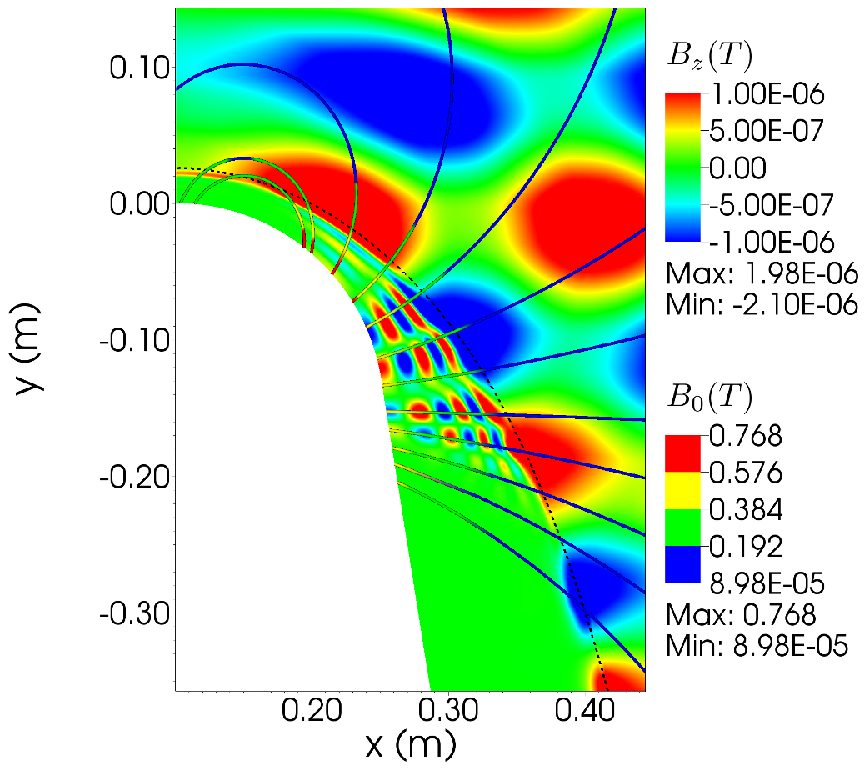}
\label{fig:lowWhistlerBzNew}}
\caption[]{EM wave propagation in whistler mode in to the plasma layer of RAM C. The imposed magnetic field is shown by the streamlines. The three components of the electric and magnetic field are shown by the contour floods of \subref{fig:lowWhistlerExNew}, \subref{fig:lowWhistlerEyNew}, \subref{fig:lowWhistlerEzNew}, \subref{fig:lowWhistlerBxNew}, \subref{fig:lowWhistlerByNew}, \subref{fig:lowWhistlerBzNew} respectively. The black dashed contour line corresponds to the electron plasma frequency of 1.6 GHz.}
\vspace{-0.2in}
\label{fig:magneticWindow}
\end{center}
\end{figure}

The magnetic field \change[]{was}{is} applied on the surface near to the nosecap using a current carrying coil of radius 0.1 m centered at (0.15, -0.15). \add[]{The coil's axis is normal to y-axis and lies in the xy plane.} The current \change[]{was}{is} $1.5\times10^{5}$ A.  In practice a permanent magnet would be used to generate the field.  The magnetic field lines colored in magnitude can be seen in all of the subplots of Fig.\ref{fig:magneticWindow}. The maximum field strength available on the RAM C surface is 0.77 T while the value is around 0.125T at the the edge of the plasma layer where significant propagation of wave occurs.  These magnetic field strengths are slightly large, however, many hypersonic vehicles of interest will actually have a considerably smaller plasma density and thus require a much weaker field to allow whistler wave propagation - this situation described can be considered an extreme case.  In addition, it's possible that using a weaker field will still allow whistler wave propagation as the evanescent waves may propagate through the outer edge of the plasma where the field is weak, then propagate as whistler waves closer to the vehicle surface. Figure \ref{fig:magneticWindow} shows the flood contours of electric and magnetic field components of the EM wave. The cutoff frequency of plasma $f_{pe}$ = 1.6 GHz is depicted by the dashed contour line. The three components of the electric and magnetic field are shown in Figs.\ref{fig:lowWhistlerExNew}-\ref{fig:lowWhistlerEzNew} and Figs.\ref{fig:lowWhistlerBxNew}-\ref{fig:lowWhistlerBzNew} respectively. It has to be noted that, the electric field contours are limited between -100 and 100 V/m, in order to improve the visibility of the whistler wave. It is clear now that the wave signal propagates through the plasma layer in the whistler mode. The additional components \add[]{E$_z$, B$_x$, and B$_y$} arising in Fig.\ref{fig:magneticWindow} when compared to the Fig.\ref{fig:blackout} are due to the circular polarization of the wave around the magnetic field lines. The circularly polarized wave propagates parallel to the magnetic field lines. Hence the angle between the wave vector and the magnetic field lines at the edge of the plasma layer plays an important role. The wave does not propagate along the field lines perpendicular to the direction of propagation. It can also be observed from the figure that the magnetic window not only allows the passage of the original electromagnetic wave, it also focuses the wave through the converging magnetic field, which is in agreement with the observations made in Ref.\cite{takechi1999rf}.

\begin{figure}
\begin{center}
\includegraphics[width=.5\textwidth]{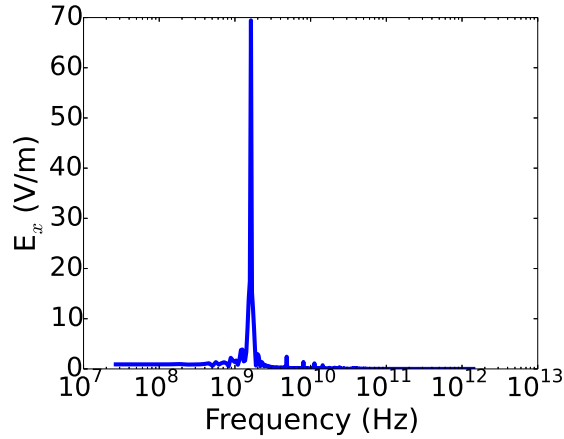}
\caption[]{Whistler wave frequency on the surface of RAM C.}
\vspace{-0.2in}
\label{fig:WhistlerFrequency}
\end{center}
\end{figure}

The ability to recover the original signal on the surface can be checked to verify that a useable signal can be obtained.  The best way to do this is to find the frequency of the signal and its energy density at the surface. The frequency of the whistler wave is obtained by taking Fourier transform of the wave history recorded on the surface of RAM C. Figure \ref{fig:WhistlerFrequency} shows the frequency of \change[]{E$_y$}{E$_x$} at (0.25555, -0.15829). The highest peak corresponds to a frequency of 1.6 GHz verifying that the original signal can be recovered at the vehicle surface. 

\begin{equation}
Q_{EM} = \frac{1}{2}\left(\epsilon_{0} \vec{E}\cdot\vec{E} + \frac{1}{\mu_{0}}\vec{B}\cdot\vec{B}\right) 
\label{eq:energyDensity}
\end{equation}

\begin{figure}
\begin{center}
\subfigure[]{
\includegraphics[width=.5\textwidth]{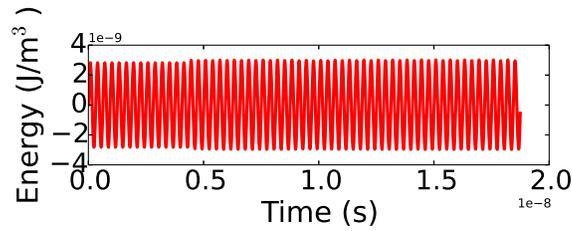}
\label{fig:energyHistoryFreeSpace}}
\subfigure[]{
\includegraphics[width=.5\textwidth]{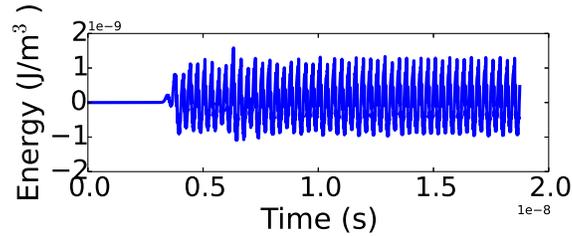}
\label{fig:energyHistoryOneFourth}}
\caption[]{The energy density recorded in the free space and on the surface of RAM C. The maximum magnetic field on the RAM C surface is 0.77 T and the magnetic field at the plasma layer's edge where the wave propagates is 0.125 T. \subref{fig:energyHistoryFreeSpace} Recorded in free space and \subref{fig:energyHistoryOneFourth} on the RAM C surface at (0.25555, -0.15829).}
\vspace{-0.2in}
\label{fig:energyHistory}
\end{center}
\end{figure}

In addition to the frequency match, the signal strength can also be calculated. The comparison of  the energy density of the signal in the free space and that on the surface gives an idea of the signal strength. The wave's energy density is computed using Eq. (\ref{eq:energyDensity}). A comparison of the electromagnetic energy density of the wave in the free space and that of the whistler wave on the surface of RAM C at (0.25555, -0.15829) is shown in Fig. \ref{fig:energyHistory}. The Fig.\ref{fig:energyHistoryFreeSpace} corresponds to the wave in the free space and the bottom subplot is for the whistler wave. The slight rise in the energy of the free space wave at t = 4.5 ns is due to the added reflected components. It can be seen from Fig.\ref{fig:energyHistoryOneFourth} that the whistler wave's energy is sufficient to be received by the antenna. The whistler wave's energy density is about 40$\%$ that of the original wave. 

\begin{figure}
\begin{center}
\includegraphics[width=.45\textwidth]{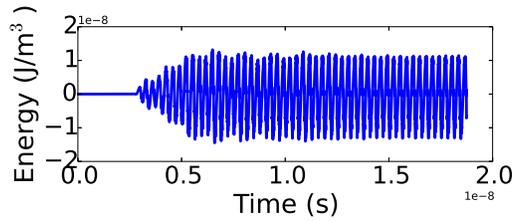}
\caption[]{The recorded energy density on the surface of RAM C at (0.258043, -0.172272). The maximum magnetic field on the RAM C surface is 3.1 T and the magnetic field at the plasma layer's edge where the wave propagates is 0.8 T.}
\vspace{-0.2in}
\label{fig:highEnergyHistory}
\end{center}
\end{figure}

Another interesting observation made during the simulations is the amplification of the whistler wave energy density with the increase of magnetic field. For instance, increasing the coil current by 4 times generates a maximum magnetic field of 3.1 T on the surface. The strength of the magnetic field at the plasma layer's edge where the whistler wave propagation occurs in this case is around 0.8 T. The wave energy density history in a magnetic widow created with the increased magnetic field is shown in Fig.\ref{fig:highEnergyHistory}. The energy density in this case is 400$\%$ of the source wave. In fact, the energy is amplified by about four times the wave's energy in the free space. As explained previously, the amplification is due to the focusing of wave along the converging field lines. This amplification could be useful in the cases where, the original signal itself is weak.

From the above feasibility analysis, it can be said that the magnetic window subjected to a realistic flight condition is capable of propagating the signal on to the vehicle's surface with energy densities ranging from 40$\%$ to 400$\%$ using magnetic fields of 0.15T and 0.8 T respectively at the edge of the plasma layer. However, the configuration can be further optimized by changing the orientation of the magnetic field lines after obtaining the plasma distributions for all the critical flight conditions in terms of angle of attack, speed and altitude. 

The designer can also test other mitigation methods using the same model described in this paper. For instance, the electrophilic fluid injection method can be tested by adding the additional reactions to the existing reactions set of the multi-species transport equations to establish the reduced plasma density. The electron acoustic wave transmission can be tested by including the multi-fluid advection terms in the analysis\cite{mudaliar2012radiation} so that the electron acoustic wave is simulated. Similarly, resonant transmission can be modeled with the equations described.

\section{Conclusions}
A procedure to model and simulate the hypersonic flow and the vehicle's communication blackout is shown. The plasma density on the RAM C vehicle showed good agreement with the reflectometer measurements from the literature. Addition of radiation losses to the reactive flow could further improve the accuracy the simulation results. The results of the Maxwell equation solver of USim are validated with the analytical solution of whistler wave propagation in one dimensional plasma layer. The Whistler mode propagation of the wave on the RAM C surface is demonstrated successfully. The frequency and the energy density of the wave signal recorded on the surface of RAM C showed a good possibility of recovering the signal propagated in whistler mode.  Although only the magnetic window \change[]{was}{is} investigated, the same plasma model together with the solver can be used to investigate many radio blackout mitigation schemes including electron acoustic wave transmission and resonant transmission.

\section*{Acknowledgments}
The authors are thankful to the financial support from AFOSR (grant numbers FA9550-12-C-0039 and FA9550-14-C-0004). The authors thank Dr. Thomas Jenkins, Dr. Ming-Chieh Lin, and Dr. David Smithe of Tech-X Corporation, for their helpful comments.

\end{document}